\documentclass[sigconf,anonymous=false]{acmart}
\AtBeginDocument{%
  \providecommand\BibTeX{{%
    \normalfont B\kern-0.5em{\scshape i\kern-0.25em b}\kern-0.8em\TeX}}
    }

\setcopyright{acmcopyright}

\copyrightyear{2023}
\acmYear{2023}
\setcopyright{acmlicensed}
\acmConference[SIGIR '23]{Proceedings of the 46th International ACM SIGIR Conference on Research and Development in Information Retrieval}{July 23--27, 2023}{Taipei, Taiwan}
\acmBooktitle{Proceedings of the 46th International ACM SIGIR Conference on Research and Development in Information Retrieval (SIGIR '23), July 23--27, 2023, Taipei, Taiwan}
\acmPrice{15.00}
\acmDOI{10.1145/3539618.3591653}
\acmISBN{978-1-4503-9408-6/23/07}

\usepackage{ulem}
\usepackage{subcaption}
\usepackage{ulem}
\usepackage{multirow}
\usepackage{graphicx}
\usepackage{algorithm, algpseudocode}

\begin{document}

\title{Continuous Input Embedding Size Search For Recommender Systems}

\author{Yunke Qu}
\affiliation{
  \institution{The University of Queensland}
  \city{Brisbane}
  \state{}
  \country{Australia}
}
\email{yunke.qu@uq.net.au}

\author{Tong Chen}
\affiliation{
  \institution{The University of Queensland}
  \city{Brisbane}
  \state{}
  \country{Australia}
}
\email{tong.chen@uq.edu.au}

\author{Xiangyu	Zhao}
\affiliation{
  \institution{City University of Hong Kong}
  \city{}
  \state{}
  \country{Hong Kong}
}
\email{xy.zhao@cityu.edu.hk}

\author{Lizhen Cui}
\affiliation{
  \institution{Shandong University}
  \city{Jinan}
  \state{}
  \country{China}
}
\email{clz@sdu.edu.cn}

\author{Kai	Zheng}
\affiliation{
  \institution{School of Computer Science and Engineering and Shenzhen Institute for Advanced Study, University of Electronic Science and Engineering of China}
  \city{Chengdu}
  \state{}
  \country{China}
}
\email{zhengkai@uestc.edu.cn}

\author{Hongzhi Yin}
\authornote{Corresponding author}
\affiliation{
  \institution{The University of Queensland}
  \city{Brisbane}
  \state{}
  \country{Australia}
}
\email{h.yin1@uq.edu.au}

\renewcommand{\shortauthors}{Yunke Qu et al.}

\begin{abstract}
Latent factor models are the most popular backbones for today’s recommender systems owing to their prominent performance. Latent factor models represent users and items as real-valued embedding vectors for pairwise similarity computation, and all embeddings are traditionally restricted to a uniform size that is relatively large (e.g., 256-dimensional). With the exponentially expanding user base and item catalog in contemporary e commerce, this design is admittedly becoming memory-inefficient. To facilitate lightweight recommendation, reinforcement learning (RL) has recently opened up opportunities for identifying varying embedding sizes for different users/items. However, challenged by search efficiency and learning an optimal RL policy, existing RL-based methods are restricted to highly discrete, predefined embedding size choices. This leads to a largely overlooked potential of introducing finer granularity into embedding sizes to obtain better recommendation effectiveness under a given memory budget. In this paper, we propose continuous input embedding size search (CIESS), a novel RL-based method that operates on a continuous search space with arbitrary embedding sizes to choose from. In CIESS, we further present an innovative random walk-based exploration strategy to allow the RL policy to efficiently explore more candidate embedding sizes and converge to a better decision. CIESS is also model-agnostic and hence generalizable to a variety of latent factor recommender systems, whilst experiments on two real-world datasets have shown state-of-the-art performance of CIESS under different memory budgets when paired with three popular recommendation models. Code is available at https://github.com/qykcq/Continuous-Input-Embedding-Size-Search-For-Recommender-Systems.
\end{abstract}



\maketitle

\section{Introduction}\label{sec:intro}
Recommender systems predict a user's preference for an item based on their previous interactions with other items \cite{zhang2019deep, wang2021survey} and have been widely applied in various e-commerce services. As arguably the most representative and powerful recommendation algorithm, latent factor models use an embedding table to map user and item IDs to dedicated vector representations (i.e., embeddings). The user and item embeddings are then fed into a pairwise similarity function (e.g., dot product or deep neural networks) to predict a user's preference for each item. However, the embedding tables can be memory-inefficient and pose challenges for storage \cite{automlrecsurvey} and deployment on personal devices \cite{10.1145/3366423.3380170, 10.1145/3447548.3467220, li2021lightweight} due to the large number of users and items in modern applications.

The root cause is that, each user/item embedding in the conventional embedding table shares the same fixed embedding size. An example from \cite{lian2020lightrec} shows that a recommender model embedding 10 million items into 256-dimensional vectors can exceed 9 GB memory consumption in a double-precision float system. As such, researchers have developed better solutions to compress the embedding table while maintaining its expressiveness. The most primitive recommender uses binary codes as an alternative to embeddings \cite{10.1145/2911451.2911502}, which was soon supplanted by another line of methods that compress fixed-sized embeddings into a lightly parameterized component. For example, codebooks have been proposed to store the embedding latent vectors efficiently in \cite{lian2020lightrec, shi2020compositional, 10.1145/3580364}. \cite{10.1145/3477495.3531775} introduced the semi-tensor product operation to tensor-train decomposition to derive an ultra-compact embedding table, and  \cite{https://doi.org/10.48550/arxiv.2010.10784} encoded users and items into hash codes and applied neural layers to learn their dense representations. However, these methods must be redesigned and retrained for different memory budgets to maintain optimal recommendation performance. 

Due to this inflexibility, there has been a new line of methods featuring reinforcement learning (RL) to automatically search for variable-size embeddings while balancing the memory consumption. For instance, in \cite{liu2020automated}, a policy network was used to select the embedding size for each user/item from a predefined set of actions.  \cite{joglekar2020neural} first discretizes the embedding table and then devised a RL-based policy network that searches the optimal embedding size configuration. RL allows us to plug in a memory cost term in the reward function to adaptively adjust the embedding sizes to achieve an ideal trade-off between space complexity and accuracy. Despite improved performance, these methods are built on a highly discrete search space with a small collection of predefined embedding sizes (e.g., only six choices in \cite{liu2020automated}. Consequently, such a narrow range of choices do not necessarily contain the optimal size for each user/item. This is likely to result in suboptimal solutions since the suggested embedding sizes may be either too small to guarantee expressiveness of important users'/items' representations, or too huge to be memory-efficient. 
 
Ideally, an embedding search paradigm should allow each user/item to have arbitrary embedding sizes, maximizing the potential of obtaining optimal performance. However, non-trivial challenges have to be addressed before we can allow for such freedom of candidate embedding sizes in the search space. Firstly, although a straightforward solution is to discretize the action space by treating every integer in the interval (e.g., $[1,256]$) as a candidate size, the policy network will be prone to suboptimal effectiveness due to the vast action space \cite{https://doi.org/10.48550/arxiv.1809.02121, ddpg}.
Secondly, such conversions will also pose challenges on training efficiency. On the one hand, the commonly used greedy exploration, i.e., iteratively selecting the action with the maximum expected reward will involve the costly ``train-and-evaluate'' cycle under every possible embedding size, which can quickly become computationally prohibitive in the recommendation setting. On the other hand, despite the possible remedies from learning parameterized functions to estimate an action's quality values (i.e., Q-values) \cite{dulac2015deep}, training and evaluating such parameterized functions still requires sufficient coverage of the embedding sizes assigned to different users/items, which again brings back the performance bottleneck. 

In light of these challenges, we propose continuous input embedding size search (CIESS), which is an RL-based algorithm that can efficiently operate in a (near\footnote{This is due to the fact that embedding sizes can only be integers.}) continuous action space for embedding size search. To enable generalization to a continuous action space, we build CIESS upon an actor-critic paradigm \cite{grondman2012survey}, specifically the twin delayed deep deterministic policy gradient (TD3) \cite{td3}, where we have designed a policy/actor network that can determine the best embedding size from an interval based on the state of each user/item. Compared with existing embedding size search counterparts \cite{joglekar2020neural,zhaok2021autoemb,liu2020automated} that only support discrete embedding size search from a small pool of actions, this is the first work to explore a large, continuous, and fine-grained RL search space of embedding sizes, thus unlocking the full potential of learning a compact recommender system with optimal recommendation performance. Furthermore, CIESS is a versatile embedding size search approach that does not hold any assumptions on the backbone recommendation model, and is compatible with a variety of latent factor recommenders that require an embedding table. 

However, given the large number of possible embedding sizes, it is unlikely the actor will always reach the states with the highest reward, introducing high variance in the estimation of the optimal Q-value. In short, when applied to continuous embedding size search, the actor in TD3 will be tasked to maximize Q-values computed by the parameterized estimator (i.e., the critic) that is hard to train and potentially erroneous, leading to inferior performance. In CIESS, we propose to explore a group of candidate embedding sizes, and select the one with the maximum Q-value in each iteration. To achieve this without greedily evaluating all possible actions, we innovatively design a \textit{random walk} mechanism in our actor-critic optimization. By performing random walks from the original action produced by the actor networks, the actors sample a small sequence of alternative actions similar to the current one. Next, the actor passes this sequence of actions to the critic for selecting the most rewarding action, which is in turn used to optimize the actor. Intuitively, this brings controlled mutations to currently the best embedding size selected, and pushes CIESS to explore better choices with higher rewards. We will empirically show that the random walk component endows the actor with improved convergence, and hence stronger utility of the resulted embedding table after compression. 

To sum up, our work entails the following contributions:
\begin{itemize}
    \item We point out that relaxing the discrete and narrow action space into a continuous one with arbitrary dimensionality choices yields better expressiveness of compressed embeddings for RL-based recommender embedding size search.
    \item We propose CIESS, a novel embedding size search method with RL. CIESS innovatively operates on a continuous interval to locate the best embedding size for each user/item, where a random walk-based actor-critic scheme is designed to guarantee optimal embedding size decisions amid the substantially enlarged action space.
    \item We conduct extensive experimental comparisons with state-of-the-art baselines paired with a variety of base recommenders, where the results have verified the advantageous efficacy of CIESS.
\end{itemize}

\section{Methodology}
\begin{figure}
    \centering
    \includegraphics[width=\linewidth]{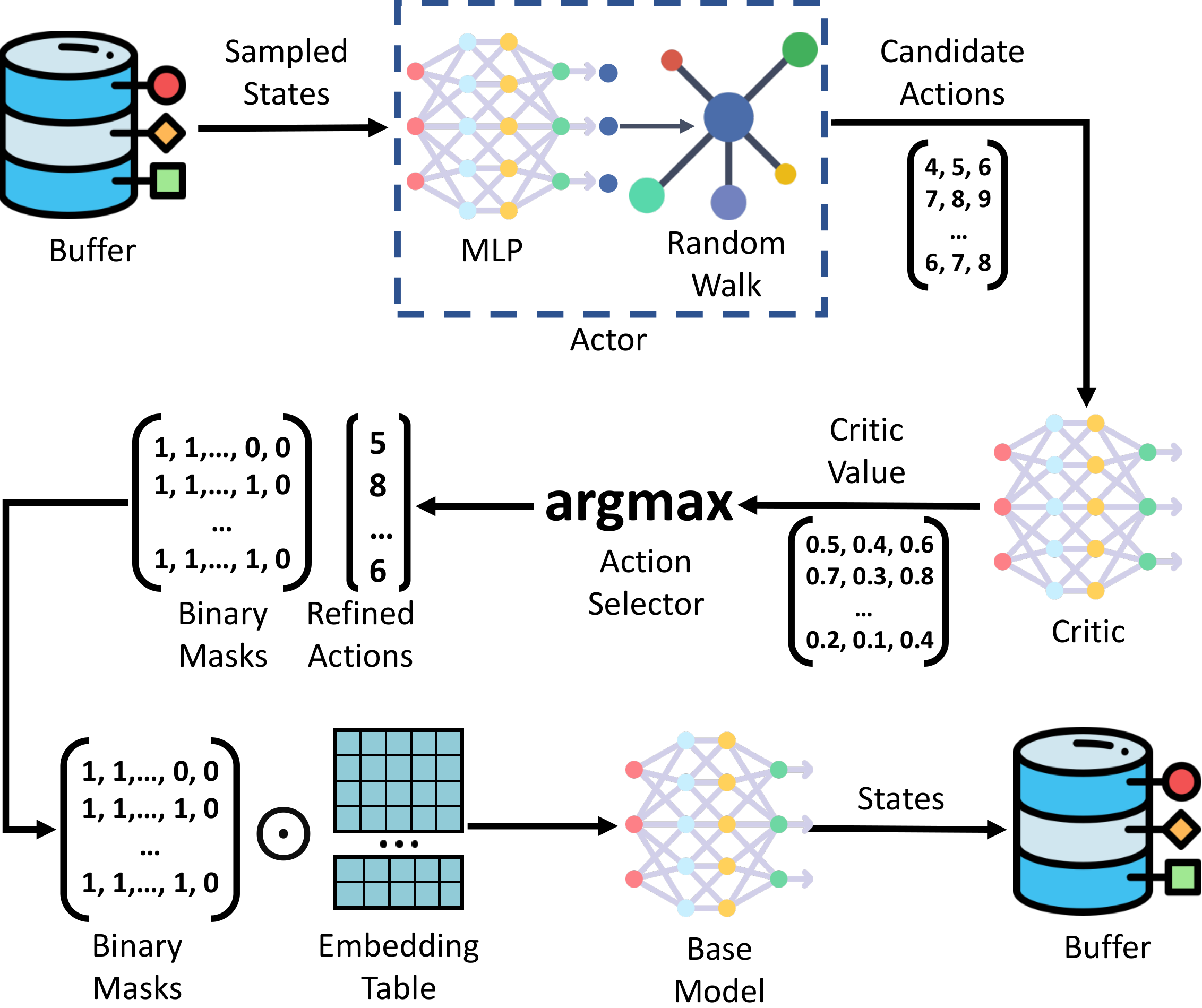} 
    \caption{An overarching view of CIESS.} \label{fig:overview}
\end{figure}

CIESS has two main components that work alternately during training: (1) a recommendation model $F_{\Theta}(\cdot)$ parameterized by $\Theta$; and (2) the RL-based search function $G_{\Phi}(\cdot)$ parameterized by $\Phi$. A schematic view of CIESS's workflow is shown in Figure~\ref{fig:overview}. In each optimization iteration of CIESS, the recommender $F_{\Theta}(\cdot)$ adjusts its user/item embedding sizes to the ones provided by the policy $G_{\Phi}(\cdot)$, then updates its parameters $\Theta$ with training samples. Afterwards, $F_{\Theta}(\cdot)$ is evaluated on a hold-out dataset, where the top-$k$ recommendation quality can be measured by common metrics such as Recall at Rank $k$ (Recall@$k$) and Normalized Discounted Cumulative Gain at Rank $k$ (NDCG@$k$). Based on the recommendation quality, the search function $G_{\Phi}(\cdot)$ will be revised, and then updates its embedding size selection for each user/item for the next iteration. In what follows, we unfold the design of CIESS.

\subsection{Base Recommender with Masked Embeddings} \label{backbonerecsys}
Let $\mathcal{U}$ and $\mathcal{V}$ be a set of users $u$ and items $v$, respectively. Their embedding vectors are stored in a real-valued embedding table $\mathbf{E}$ with the dimensionality of $(|\mathcal{U}|+|\mathcal{V}|) \times d_{max}$. 
It can be viewed as the vertical concatenation of all user and item embeddings $[\mathbf{e}_{u_1};...;\mathbf{e}_{u_{|\mathcal{U}|}};\mathbf{e}_{v_1};...;\mathbf{e}_{v_{|\mathcal{V}|}}]$, where $d_{max}$ is the initial embedding dimension of all users/items in the full-size embedding table. In other words, $d_{max}$ is also the maximum embedding size in the search space.

By performing embedding look-ups, we can map each user/item ID to a real-valued embedding vector $\mathbf{e}_{u}$ and $\mathbf{e}_{v}$. To enable adjustable embedding sizes, we introduce a binary mask $\mathbf{M}\in \{0,1\}^{(|\mathcal{U}|+|\mathcal{V}|) \times d_{max}}$, which is applied to $\mathbf{E}$ during the embedding look-up:
\begin{equation}
    \mathbf{e}_{n} = \text{Lookup}(\mathbf{E} \odot \mathbf{M}, n), \;\;\;\; n\in\mathcal{U}\cup \mathcal{V},
\end{equation}
where $\odot$ is the element-wise multiplication and $n$ is the ID of a user/item. \textbf{For simplicity, we use} $n\in\mathcal{U}\cup \mathcal{V}$ \textbf{to index either a user/item when there is no ambiguity}. Notably, $\mathbf{M}$ is dynamically updated according to the current policy to control the usable dimensions of each embedding vector. Given an automatically learned embedding size $d_n$ for a specific user/item, the $s$-th element of its corresponding mask vector $\mathbf{m}_n$ is defined as:
\begin{equation}\label{eq:mask}
    \mathbf{m}_{n(s)} = 
    \begin{cases}
        1 & \text{for }1 \le s \le d_n\\    
        0 & \text{for }d_n < s < d_{max}
    \end{cases}, \;\;\;\; n\in\mathcal{U}\cup \mathcal{V}.
\end{equation}
With the mask $\mathbf{M}$, for each user/item, we can retain the first $d_n$ elements of its full embedding while setting all succeeding dimensions to $0$. It is worth noting that, performing embedding sparsification by masking unwanted dimensions with zeros is a commonly adopted approach in lightweight recommender systems \cite{liu2021learnable, optemb}, as the resulting embedding table can take advantage of the latest sparse matrix storage techniques \cite{virtanen2020scipy,sedaghati2015automatic} that bring negligible cost for storing zero-valued entries.

On obtaining the sparsified embeddings of both users and items, the recommendation model $F_{\Theta}(\cdot)$ will output a preference score $\hat{y}_{uv}$ denoting the pairwise user-item affinity:
\begin{equation}
    \hat{y}_{uv} = F_{\Theta}(\mathbf{e}_u,\mathbf{e}_{v}), \;\;\;\; u\in\mathcal{U}, v\in \mathcal{V},
\end{equation}
where the choice of $F_{\Theta}(\cdot)$ can be arbitrary, as long as it supports such pairwise similarity computation with user and item embeddings. 

\textbf{Objective w.r.t. Embedding Size Search.} For optimizing the recommender, we adopt the well-established Bayesian Personalized Ranking (BPR) Loss \cite{rendle_bpr_2012}:
\begin{equation} 
   \mathcal{L}_{BPR} = \sum_{(u, v, v') \in D} - \ln \sigma (\hat{y}_{uv} - \hat{y}_{uv'}) + \gamma\|\Theta\|_2^2,
\end{equation}
where $\mathcal{D}$ denotes the training dataset, $(u, v, v')$ denotes the user $u$ prefers item $v$ over item $v'$, and $\hat{y}_{uv}$ and $\hat{y}_{uv'}$ are the predicted preferences that the user $u$ has for items $v$ and $v'$. The second term is the $L2$ regularization weighted by $\gamma$ for overfitting prevention. As we are interested in performing embedding size search for each individual user and item under a given memory budget, we define the overall objective as follows:
\begin{equation}\label{eq:objective}
    \min_{\Theta,\Phi} \mathcal{L}_{BPR} \;\;\;\text{s.t.}\;1- \frac{\| \mathbf{M}\|_{1,1}}{(|\mathcal{U}|+|\mathcal{V}|) \times d_{max}} \ge c,
\end{equation}
where our target sparsity ratio $c$ ($0< c <1$) specifies the percentage of parameters to be pruned (i.e., zeroed out in our case) in the final sparse embedding matrix $\mathbf{E} \odot \mathbf{M}$.   

\subsection{Continuous Embedding Size Search with Reinforcement Learning}
Now that the base recommender can accommodate varying embedding sizes via masked sparsification, we start to search for the optimal embedding sizes with $G_{\Phi}(\cdot)$. In order to efficiently learn a quality embedding size search policy from a continuous action space, we hereby introduce our solution in an RL setting by presenting our design of the environment, state, action, reward, actor and critic.

\subsubsection{Environment}
As discussed in Section \ref{backbonerecsys}, the base recommender model allows for adjustable embedding sizes. During the optimization process, the environment receives the action (i.e., embedding sizes for all user/items), provides feedback (i.e., reward) on both the memory cost and recommendation performance, and update its state for the subsequent action prediction.  

\subsubsection{State} \label{subsubsec:state}
The state $s$ is the input to the policy network (i.e., actor in CIESS) that drives the decision-making on each user/item-specific embedding size. 
\cite{zhaok2021autoemb} shows that the popularity (i.e., the number of interactions) and the current embedding size decision of the user/item are effective in providing the policy network the context for subsequent search. Our method inherits this design, with an additional quality indicator $a$ that records the recommendation accuracy fluctuation under the current policy:
\begin{equation}\label{eq:state}
    s_{n} = (\frac{f_{n} - f_{min}}{f_{max} - f_{min}}, \frac{d_{n} - d_{min}}{d_{max} - d_{min}}, q_{n}), \;\;\;n\in\mathcal{U}\cup\mathcal{V},
\end{equation}
where $f_n$ is the popularity of the user/item, normalized by the corresponding maximum/minimum frequency among all users $f_{max}=\max_{u\in\mathcal{U}}(f_u)$/$f_{min}=\min_{u\in\mathcal{U}}(f_u)$ or items $f_{max}=\max_{v\in\mathcal{V}}(f_v)$/$f_{min}=\min_{v\in\mathcal{V}}(f_v)$, which are observed from training data. $d_n$ is the current embedding size allocated to a user/item, and $q_n$ quantifies the changes in the recommendation quality when the embedding size decreases from $d_{max}$ to the current $d_n$. Compared with \cite{zhaok2021autoemb}, incorporating this quality indicator into the state is able to help trace the impact from the most recent action (i.e., embedding size) to the recommendation effectiveness, which can encourage the policy network to better balance the memory cost and performance with the embedding sizes selected from a vast, continuous action space.

For $q_n$, we define it as the ratio between the current ranking quality under the currently selected embedding size $d_n$ and the ranking quality under the initial/maximal embedding size $d_{max}$:
\begin{equation}\label{eq:q_compute}
    q_n = \min(\frac{\text{eval}(\mathbf{e}_n|\mathbf{E} \odot \mathbf{M})}{\text{eval}(\mathbf{e}_n|\mathbf{E})},\,1),\;\;\;\;n\in\mathcal{U}\cup\mathcal{V},
\end{equation}
where $\text{eval}(\cdot)$ evaluates the recommendation performance w.r.t. a user/item embedding $\mathbf{e}_n$ drawn from the specified embedding table. With the $\min(\cdot)$ operator, we restrict $q_n\in [0,1]$. The denominator $\text{eval}(\mathbf{e}_n|\mathbf{E})$ can be precomputed with the fully trained based recommender and reused throughout the search process. Instead of using the raw recommendation accuracy measure $\text{eval}(\mathbf{e}_n|\mathbf{E} \odot \mathbf{M})$, we use the ratio format in Eq.(\ref{eq:q_compute}) to indicate the fluctuation (mostly reduction) of recommendation quality when the embeddings are compressed.  
This is due to the fact that some users and items are inherently trivial or hard to rank, e.g., long-tail users tend to have a relatively lower recommendation accuracy even with full embeddings, and adjusting their embedding sizes will not significantly affect the accuracy obtained. In such cases, using the performance ratio prevents misleading signals and amplifies the reward for those users/items.

Since Recall@$k$ and NDCG@$k$ are common metrics \cite{he2020lightgcn, krichene2022sampled} for recommendation evaluation, we implement an ensemble of both Recall and NDCG scores under different $k$ values for $\text{eval}(\cdot)$:
\begin{equation}\label{eq:eval_user}
    \text{eval}(\mathbf{e}_u|\cdot) = \frac{\sum_{k \in \mathcal{K}} \text{Recall@}k_u + \text{NDCG}@k_u}{2|\mathcal{K}|}, \;\;\;\; u\in\mathcal{U},
\end{equation}
where the choices of $k$ in our paper are $\mathcal{K}=\{5,10,20\}$, and Recall@$k_u$ and NDCG@$k_u$ denote the evaluation scores w.r.t. user $u \in\mathcal{U}$. 

Note that, Eq.(\ref{eq:eval_user}) is only applicable to user embeddings, as recommendation metrics are essentially user-oriented. Hence, for item $v\in\mathcal{V}$ and its embedding $\mathbf{e}_v$, we identify its interacted users $\mathcal{U}_v$ in the training set, then take the average of all $\text{eval}(\mathbf{e}_u|\cdot)$ scores for $u\in \mathcal{U}_v$:
\begin{equation}
    \text{eval}(\mathbf{e}_v|\cdot) = \frac{1}{|\mathcal{U}_v|} \sum_{u \in \mathcal{U}_v}\text{eval}(\mathbf{e}_u|\cdot), \;\;\;\; v\in\mathcal{V},
\end{equation}
where $\mathbf{e}_u$ and $\mathbf{e}_v$ are drawn from the same embedding table.

\subsubsection{Reward} \label{subsubsec: reward}
The reward $r$ is the feedback to the current policy to guide subsequent embedding size adjustments. Given our goal stated in Eq.(\ref{eq:objective}), the reward should reflect a blend of recommendation quality and space complexity. Therefore, we design the following pointwise reward for a user/item on their current embedding size $d_u$/$d_v$ selected:
\begin{equation}\label{eq:reward}
    r_{n} = q_{n} - \lambda (\frac{d_{n}}{d_{max}})^2 ,\;\;n\in \mathcal{U}\cup\mathcal{V}, 
\end{equation}
where the first term is the ranking quality defined in Eq.(\ref{eq:q_compute}), and the second term weighted by the scaling hyperparameter $\lambda$ measures the memory cost of the embedding size chosen. The squared form in the memory cost magnifies the reward gain during early pruning stages (e.g., when $\frac{d_{n}}{d_{max}}$ drops from $1$ to $0.9$) and stabilizes that at later stages (e.g., when $\frac{d_{n}}{d_{max}}$ drops from $0.2$ to $0.1$). As such, we can stimulate a sharper embedding size reduction initially to quickly approach the target sparsity ratio $c$, and then encourage fine-grained action selection when optimizing towards the balance between performance and space.

\subsubsection{Action}\label{sec:action}
At every iteration $i$, the policy network in the search function $G_{\Phi}$ predicts an action, i.e., dimension $d_n^i\in \mathbb{N}_{>0}$ from interval $[1,d_{max}]$ given a user/item state $s_n^i$. The action is the embedding size for the corresponding user/item. The recommender $F_{\Theta}(\cdot)$ takes this action by altering the embedding sizes of users/items (i.e., updating $\mathbf{M}$ in practice), and yields a reward $r^i_n$ along with a subsequent state $s_n^{i+1}$. The tuple $(s^i_n, d^i_n, r^i_n, s^{i+1}_n)$ is defined as a \textit{transition}, which are stored in a replay buffer $\mathcal{B}$ for subsequent training.

\subsubsection{Actor and Critic}
We adopt an actor-critic paradigm for RL, which provides a better convergence guarantee and lower variance compared with pure policy search methods \cite{grondman2012survey}. Unlike policy-based RL that fails to generalize to continuous action space and cannot extend to unseen actions \cite{dulac2015deep}, we adopt a continuous RL backbone, namely TD3 \cite{td3}. In CIESS, we have two actor networks $\mu_{\mathcal{U}}(\cdot)$ and ${\mu}_{\mathcal{V}}(\cdot)$ respectively built for the user and item sets. The rationale for having two separate actors is to better accommodate the distributional difference between user and item states, and more concretely, the popularity $f_n/f_{max}$ and recommendation quality indicator $q_n$. These two actors share the same architecture and only differ in their parameterization. The action/embedding size at the $i$-th iteration is computed using the actor network given the current user/item state:
\begin{align}\label{eq:actor}
    \hat{d}_{u}^i &= \mu_{\mathcal{U}}(s_u^i) + \epsilon, \;\;\; u \in \mathcal{U} \\  \nonumber
    \hat{d}_{v}^i &= \mu_{\mathcal{V}}(s_v^i) + \epsilon, \;\;\; v \in \mathcal{V}
\end{align}
where we add a small noise $\epsilon \sim \mathcal{N}(0, \sigma)$ from Gaussian distribution with standard deviation $\sigma$. This is to introduce a small amount of variation to the computed action, hence allowing additional explorations on more possible embedding sizes for the policy. 

\begin{figure}
    \centering
    \includegraphics[width=\linewidth]{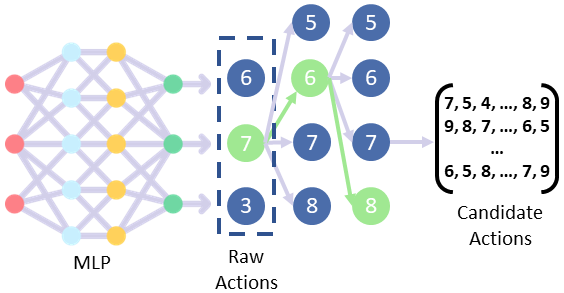}
    \caption{The random walk-based embedding size exploration, where raw actions are $\hat{d}^i_n$ predicted by the actor.} \label{fig:rw}
\end{figure}

Correspondingly, we build user- and item-specific critic networks to learn to approximate the quality (i.e., Q-value) of an action $d^i_n$ taken at state $s^i_n$, denoted as $Q_{\mathcal{U}}(s_u^i, d_u^i)$ and $Q_{\mathcal{V}}(s_v^i, d_v^i)$, respectively. Instead of the traditional value-based RL that additionally requires learning a value function \cite{joglekar2020neural}, the critic networks in the actor-critic framework are trained via tuples stored in the replay buffer $\mathcal{B}$ to map the actor's interaction with the environment to the Q-value, which directs the actors to make policy updates by estimating the quality of the actions taken. 

\textbf{Action Exploration with Random Walk.} As discussed in Section \ref{sec:intro}, such value-based RL optimization heavily relies on learning an accurate critic, which can hardly be guaranteed in continuous embedding size search tasks with numerous actions and states. Although TD3 additionally trains an independent critic network (two additional user/item critic networks in our case) and uses the minimum of their outputs as the predicted Q-value to lower the estimation variance, such a conservative strategy may result in underestimation of the Q-value and eventually suboptimal performance \cite{liu2023actor}. Therefore, a common practice is to adopt exploration strategies on possible actions. A typical example is the greedy strategy that always selects the action with highest Q-value \cite{dulac2015deep}, which can be extremely inefficient to obtain in our large action space. Hence, in addition to the exploration noise in Eq.(\ref{eq:actor}), we further propose a sample-efficient action exploration strategy based on random walk, as shown in Figure \ref{fig:rw}. To avoid evaluating all possible actions at each state as in the greedy strategy, we only compute Q-values on a sequence of actions mutated from the computed $\hat{d}^i_n$, where the action returning the highest Q-value will be selected. The rationale is, for a given user/item, if we keep fine-tuning its embedding size based on currently the best one $\hat{d}^i_n$ predicted by the actor, there will be a higher chance that we can obtain a more performant embedding size choice. Specifically, to efficiently identify the sequence of actions for exploration, we construct a graph of actions where each action $d$ is connected to a set of neighbor actions $\mathcal{A}_d$. Since each action is an integer-valued embedding size, we can easily measure the distance between any two actions by taking the absolute value of their arithmetic difference. For each action $d$, the distance to any of its neighbors is below a predefined threshold $t$:
\begin{equation}
    \forall d' \in \mathcal{A}_d: \;\;\;|d' - d| \leq t. 
\end{equation}
From each action $d$ in the constructed graph, the probability of reaching action $d' \in \mathcal{A}_{d}$ during random walk is:
\begin{equation}
    p(d\rightarrow d') = \frac{|d' - d|}{\sum_{d' \in \mathcal{A}_{d}} |d' - d|}.
\end{equation}
Then, starting from $\hat{d}^i_n$, we perform random walk (with replacement) to obtain a series of actions $\mathcal{Z}_{\hat{d}^i_n}$, where we specify $|\mathcal{Z}_{\hat{d}^i_n}|$ to be relatively small to balance efficiency and exploration thoroughness. After obtaining $\mathcal{Z}_{\hat{d}^i_n}$, we evaluate each action in $\mathcal{Z}_{\hat{d}^i_n}$ with the critic network $Q_\mathcal{U}(\cdot)$/$Q_\mathcal{V}(\cdot)$ and greedily select the final action $d^{i}_n$ for iteration $i$:
\begin{align}\label{eq:final_action}
	d^{i}_u &= \text{argmax}_{d' \in \mathcal{Z}_{\hat{d}^i_u}} Q_\mathcal{U}(s^i_u,d^i_u),\;\;\;\; u \in \mathcal{U},\nonumber\\
	d^{i}_v &= \text{argmax}_{d' \in \mathcal{Z}_{\hat{d}^i_v}} Q_\mathcal{V}(s^i_v,d^i_v),\;\;\;\; v \in \mathcal{V},
\end{align} 
which will be utilized to optimize both the actor and critic networks. 

\subsection{Selective Retraining for Sparsified Embeddings}\label{sec:selec_retrain}
We put together a pseudo code for CIESS in Algorithm \ref{alg:cap}. The embedding size search policy is trained with RL for $M$ episodes, where each episode iterates for $T$ times. In each iteration, CIESS performs random walk-based action exploration and decides the embedding sizes for all users and items  (lines 6-9), trains the recommender $F_{\Theta}(\cdot)$ with the embedding sizes selected to obtain the instant reward $r^i_n$ and next iteration's state $s^{i+1}_n$ (lines 10-11). The transition described in Section \ref{sec:action} is appended to the replay buffer $\mathcal{B}$ to facilitate optimization of the critic and actor networks in the search function (lines 12-14). We omit the twin network-based optimization process in TD3 and set a pointer to the original paper \cite{td3} for finer details. Notably, as a fresh recommender $F_{\Theta}(\cdot)$ needs to be trained from scratch for every iteration's embedding size decision, we restrict the training process (line 10) to only run for a specified number of epochs (5 in our case) over $\mathcal{D}$ to ensure efficiency. By fixing all hyperparameters and only varying $\mathbf{M}$ for $F_{\Theta}(\cdot)$, this offers us sufficient confidence in comparing different embedding sizes' performance potential without an excessively time-consuming process. We hereby introduce how to obtain a fully trained recommender under a specific memory budget, which corresponds to line 17 in Algorithm~\ref{alg:cap}.

\begin{algorithm}[t]
\caption{CIESS}
\label{alg:cap}
\begin{algorithmic}[1]
\State Initialize the RL-based search function $G_{\Phi}(\cdot)$;
\State Initialize replay buffer $\mathcal{B}$;
\For{$episode =1,\cdots,M$}
\State Compute initial state $s^{0}_n$ w.r.t. $d_{max}$ for $n\in \mathcal{U}\cup \mathcal{V}$;
    \For{$i = 1,\cdots,T$}
    	\State /*\;Each iteration applies to all $n\in \mathcal{U}\cup \mathcal{V}$\;*/
        \State Initialize base recommender $F_{\Theta}(\cdot)$;
        \State Perform random walk from $\hat{d}_n^i \leftarrow$ Eq.(\ref{eq:actor}); 
        \State Obtain $d^i_n \leftarrow$ Eq.(\ref{eq:final_action}) and update $\mathbf{M} \leftarrow$ Eq.(\ref{eq:mask});     
        \State Update $F_{\Theta}(\cdot)$ w.r.t. Eq.(\ref{eq:objective}) and $\mathbf{E}\odot\mathbf{M}$;
        \State Evaluate $F_{\Theta}(\cdot)\!$ to obtain $r_{n}^i \!\!\leftarrow\!$ Eq.(\ref{eq:reward}), $s_{n}^{i+1}\!\!\leftarrow\!$ Eq.(\ref{eq:state});
        \State Update buffer $\mathcal{B} \leftarrow \mathcal{B}\cup (s_n^i, d_n^i, r_n^i, s_{n}^{i+1})$; 
        \State Draw a batch of transitions from $\mathcal{B}$;
        \State Update $Q_\mathcal{U}(\cdot)$, $\!Q_\mathcal{V}(\cdot)$, $\!\mu_\mathcal{U}(\cdot)$, $\!\mu_\mathcal{V}(\cdot)$ with TD3 \cite{td3};
    \EndFor
\EndFor
\State Perform selective retraining and obtain $\textbf{M}^* \leftarrow$ Eq.(\ref{eq:selec_retrain}).
\end{algorithmic}
\end{algorithm}

As stated in Eq.(\ref{eq:objective}), $c$ essentially defines the sparsity of the pruned embedding table $\mathbf{E}\odot\mathbf{M}$, e.g., $c=0.9$ means only $10\%$ of the parameters in $\mathbf{E}$ are kept. In the RL-based search stage, the policy adjusts the embedding size of each user/item (hence the embedding sparsity) until the reward is maximized. By adjusting $\lambda$ in the reward, we can let the search function $G_{\Phi}(\cdot)$ to either emphasis recommendation accuracy or memory efficiency, where the latter is preferred and adopted in this paper as we prioritize high compression rates. As such, $G_{\Phi}(\cdot)$ will be able to derive a decreasing embedding size for each user/item in the pursuit of maximum reward. 
However, recall that when obtaining the reward w.r.t. the embedding mask $\mathbf{M}$ computed in each iteration, CIESS does not update the base recommender $F_{\Theta}(\cdot)$ until full convergence. To make the final decision more robust, we do not rely on a single embedding size decision, and instead maintain a set of candidate embedding mask matrices $\mathcal{M}_{c}$ with the top-$l$ highest performance measured by $q_n$ during the search stage, constrained by $c$:
\begin{equation}
    \forall \mathbf{M}\in \mathcal{M}_{c}: \;\;\;\; 1- \frac{\| \mathbf{M}\|_{1,1}}{(|\mathcal{U}|+|\mathcal{V}|) \times d_{max}} \ge c
\end{equation}
Then, in the parameter retraining stage, we retrain the randomly initialized recommender model $F_{\Theta}(\cdot)$ for each $\mathbf{M} \in \mathcal{M}_{c}$ till convergence. Afterwards, we select matrix $\textbf{M}^*$ that yields the highest recommendation quality as the final solution:
\begin{equation}\label{eq:selec_retrain}
    \textbf{M}^* = \text{argmax}_{\mathbf{M} \in \mathcal{M}_{c}} \, \overline{q} ,
\end{equation}
where $\overline{q}$ denotes the mean of all $q_n$ for $n\in \mathcal{U}\cup\mathcal{V}$. If a lower target sparsity $c' < c$ is needed, we can further expand this selective retraining scheme to $l$ best-performing masks $\mathcal{M}_{c'}$ w.r.t. $c'$,  thus finding the optimal embedding sizes for different memory budgets in one shot.

\section{Experiments}
We detail our experimental analysis on the performance of CIESS in this section.

\subsection{Base Recommenders and Comparative Methods}
CIESS can be paired with various base recommender models that utilize embedding-based representation learning. To thoroughly validate our method's versatility and generalizability across different base recommenders, we leverage three widely used recommenders to serve as the base recommender $F_{\Theta}(\cdot)$, namely neural collaborative filtering (NCF) \cite{he2017neural}, neural graph collaborative filtering (NGCF) \cite{wang2019neural}, and light graph convolution network (LightGCN) \cite{he2020lightgcn}, where we inherit the optimal settings reported in their original work and only substitute the embedding table into a sparsified one. 

We compete against the following embedding size search algorithms, which are all model-agnostic:
\begin{itemize}
    \item PEP \cite{liu2021learnable}: It learns soft pruning thresholds with a reparameterization trick to achieve sparsified embeddings.
    \item ESAPN \cite{liu2020automated}: It is an RL-based method that automatically searches the embedding size for each user and item from a discrete action space.
    \item OptEmbed \cite{optemb}: It trains a supernet to learn field-wise thresholds, and then uses evolutionary search to derive the optimal embedding size for each field.
    \item Equal Sizes (ES): Its embedding sizes are equal across all users/items and remain fixed.
    \item Mixed and Random (MR): Its embedding sizes are sampled from a uniform distribution and remain fixed.
\end{itemize}

\begin{table*}[t!]
\resizebox{\textwidth}{!}{%
\begin{tabular}{|cccccccccccccccc}
\multicolumn{1}{c|}{} & \multicolumn{5}{c|}{LightGCN} & \multicolumn{5}{c|}{NGCF} & \multicolumn{5}{c|}{NCF} \\ \hline
\multicolumn{1}{c|}{Method} & Sparsity & R@5 & R@20 & N@5 & \multicolumn{1}{c|}{N@20} & Sparsity & R@5 & R@20 & N@5 & \multicolumn{1}{c|}{N@20} & Sparsity & R@5 & R@20 & N@5 & \multicolumn{1}{c|}{N@20} \\ \hline
\multicolumn{1}{c|}{ESAPN} & 72\% & 0.0912 & 0.2422 & 0.4771 & \multicolumn{1}{c|}{0.4178} & 85\% & 0.0856 & 0.2276 & 0.4285 & \multicolumn{1}{c|}{0.3829} & 72\% & 0.0845 & 0.2283 & 0.4454 & \multicolumn{1}{c|}{0.3822} \\ \hline
\multicolumn{1}{c|}{OptEmbed} & 83\% & 0.0745 & 0.1994 & 0.4257 & \multicolumn{1}{c|}{0.3650} & 80\% & 0.0717 & 0.2038 & 0.4045 & \multicolumn{1}{c|}{0.3622} & 85\% & 0.0458 & 0.1352 & 0.2960 & \multicolumn{1}{c|}{0.2573} \\ \hline
\multicolumn{1}{c|}{PEP} & \multirow{4}{*}{80\%} & 0.0771 & 0.2098 & 0.4346 & \multicolumn{1}{c|}{0.3778} & \multirow{4}{*}{80\%} & \textbf{0.0806} & 0.2138 & 0.4382 & \multicolumn{1}{c|}{0.3798} & \multirow{4}{*}{80\%} & 0.0725 & 0.2045 & 0.4054 & \multicolumn{1}{c|}{0.3603} \\
\multicolumn{1}{c|}{ES} &  & 0.0825 & 0.2248 & 0.4536 & \multicolumn{1}{c|}{0.3969} &  & 0.0803 & \textbf{0.2232} & 0.4281 & \multicolumn{1}{c|}{\textbf{0.3858}} &  & 0.0779 & 0.2080 & 0.4121 & \multicolumn{1}{c|}{0.3682} \\
\multicolumn{1}{c|}{MR} &  & 0.0737 & 0.2004 & 0.4211 & \multicolumn{1}{c|}{0.3648} &  & 0.0748 & 0.2053 & 0.4202 & \multicolumn{1}{c|}{0.3693} &  & 0.0692 & 0.1914 & 0.3787 & \multicolumn{1}{c|}{0.3374} \\
\multicolumn{1}{c|}{CIESS} &  & \textbf{0.0920} & \textbf{0.2436} & \textbf{0.4854} & \multicolumn{1}{c|}{\textbf{0.4257}} &  & 0.0800 & 0.2148 & \textbf{0.4442} & \multicolumn{1}{c|}{0.3854} &  & \textbf{0.0792} & \textbf{0.2221} & \textbf{0.4233} & \multicolumn{1}{c|}{\textbf{0.3828}} \\ \hline
\multicolumn{1}{c|}{PEP} & \multirow{4}{*}{90\%} & 0.0733 & 0.1988 & 0.4237 & \multicolumn{1}{c|}{0.3641} & \multirow{4}{*}{90\%} & \textbf{0.0794} & \textbf{0.2092} & 0.4309 & \multicolumn{1}{c|}{0.3727} & \multirow{4}{*}{90\%} & 0.0723 & 0.2033 & 0.4053 & \multicolumn{1}{c|}{0.3591} \\
\multicolumn{1}{c|}{ES} &  & 0.0778 & 0.2105 & 0.4376 & \multicolumn{1}{c|}{0.3787} &  & 0.0747 & 0.2069 & 0.4149 & \multicolumn{1}{c|}{0.3679} &  & 0.0694 & 0.1973 & 0.4011 & \multicolumn{1}{c|}{0.3546} \\
\multicolumn{1}{c|}{MR} &  & 0.0663 & 0.1802 & 0.3948 & \multicolumn{1}{c|}{0.3383} &  & 0.0745 & 0.2056 & 0.4122 & \multicolumn{1}{c|}{0.3660} &  & 0.0605 & 0.1703 & 0.3523 & \multicolumn{1}{c|}{0.3127} \\
\multicolumn{1}{c|}{CIESS} &  & \textbf{0.0846} & \textbf{0.2248} & \textbf{0.4631} & \multicolumn{1}{c|}{\textbf{0.4023}} &  & 0.0782 & 0.2082 & \textbf{0.4385} & \multicolumn{1}{c|}{\textbf{0.3779}} &  & \textbf{0.0759} & \textbf{0.2131} & \textbf{0.4216} & \multicolumn{1}{c|}{\textbf{0.3750}} \\ \hline
\multicolumn{1}{c|}{PEP} & \multirow{4}{*}{95\%} & 0.0659 & 0.1789 & 0.3876 & \multicolumn{1}{c|}{0.3315} & \multirow{4}{*}{95\%} & 0.0762 & 0.2020 & 0.4248 & \multicolumn{1}{c|}{0.3618} & \multirow{4}{*}{95\%} & 0.0697 & 0.1955 & 0.3864 & \multicolumn{1}{c|}{0.3379} \\
\multicolumn{1}{c|}{ES} &  & 0.0646 & 0.1752 & 0.3847 & \multicolumn{1}{c|}{0.3272} &  & 0.0736 & 0.1977 & 0.4134 & \multicolumn{1}{c|}{0.3595} &  & 0.0666 & 0.1841 & 0.3881 & \multicolumn{1}{c|}{0.3358} \\
\multicolumn{1}{c|}{MR} &  & 0.0594 & 0.1634 & 0.3657 & \multicolumn{1}{c|}{0.3119} &  & 0.0728 & 0.1958 & 0.4152 & \multicolumn{1}{c|}{0.3593} &  & 0.0595 & 0.1637 & 0.3575 & \multicolumn{1}{c|}{0.3072} \\
\multicolumn{1}{c|}{CIESS} &  & \textbf{0.0744} & \textbf{0.2009} & \textbf{0.4264} & \multicolumn{1}{c|}{\textbf{0.3664}} &  & \textbf{0.0774} & \textbf{0.2055} & \textbf{0.4333} & \multicolumn{1}{c|}{\textbf{0.3727}} &  & \textbf{0.0707} & \textbf{0.1982} & \textbf{0.4082} & \multicolumn{1}{c|}{\textbf{0.3587}} \\ \hline
\multicolumn{16}{c}{(a) Results on MovieLens-1M} \\ \hline
\multicolumn{1}{c|}{ESAPN} & 76\% & 0.0257 & 0.0752 & 0.0596 & \multicolumn{1}{c|}{0.0699} & 73\% & 0.0124 & 0.0364 & 0.0232 & \multicolumn{1}{c|}{0.0297} & 74\% & 0.0154 & 0.0448 & 0.0382 & \multicolumn{1}{c|}{0.0425} \\ \hline
\multicolumn{1}{c|}{OptEmbed} & 80\% & 0.0183 & 0.0534 & 0.0424 & \multicolumn{1}{c|}{0.0489} & 67\% & 0.0133 & 0.0412 & 0.0301 & \multicolumn{1}{c|}{0.0369} & 77\% & 0.0076 & 0.0238 & 0.0161 & \multicolumn{1}{c|}{0.0203} \\ \hline
\multicolumn{1}{c|}{PEP} & \multirow{4}{*}{80\%} & 0.0253 & 0.0723 & 0.0605 & \multicolumn{1}{c|}{0.0682} & \multirow{4}{*}{80\%} & 0.0086 & 0.0275 & 0.0167 & \multicolumn{1}{c|}{0.0223} & \multirow{4}{*}{80\%} & 0.0130 & 0.0406 & 0.0236 & \multicolumn{1}{c|}{0.0327} \\
\multicolumn{1}{c|}{ES} &  & 0.0289 & 0.0822 & 0.0665 & \multicolumn{1}{c|}{0.0758} &  & 0.0223 & 0.0665 & 0.0517 & \multicolumn{1}{c|}{0.0603} &  & 0.0131 & 0.0465 & 0.0226 & \multicolumn{1}{c|}{0.0351} \\
\multicolumn{1}{c|}{MR} &  & 0.0253 & 0.0783 & 0.0588 & \multicolumn{1}{c|}{0.0698} &  & 0.0212 & 0.0618 & 0.0499 & \multicolumn{1}{c|}{0.0573} &  & 0.0101 & 0.0342 & 0.0183 & \multicolumn{1}{c|}{0.0268} \\
\multicolumn{1}{c|}{CIESS} &  & \textbf{0.0292} & \textbf{0.0839} & \textbf{0.0692} & \multicolumn{1}{c|}{\textbf{0.0783}} &  & \textbf{0.0233} & \textbf{0.0701} & \textbf{0.0566} & \multicolumn{1}{c|}{\textbf{0.0652}} &  & \textbf{0.0175} & \textbf{0.0533} & \textbf{0.0377} & \multicolumn{1}{c|}{\textbf{0.0474}} \\ \hline
\multicolumn{1}{c|}{PEP} & \multirow{4}{*}{90\%} & 0.0224 & 0.0657 & 0.0531 & \multicolumn{1}{c|}{0.0610} & \multirow{4}{*}{90\%} & 0.0080 & 0.0276 & 0.0153 & \multicolumn{1}{c|}{0.0215} & \multirow{4}{*}{90\%} & 0.0139 & 0.0427 & 0.0242 & \multicolumn{1}{c|}{0.0335} \\
\multicolumn{1}{c|}{ES} &  & 0.0230 & 0.0722 & 0.0544 & \multicolumn{1}{c|}{0.0648} &  & 0.0205 & 0.0622 & 0.0528 & \multicolumn{1}{c|}{0.0592} &  & 0.0119 & 0.0416 & 0.0201 & \multicolumn{1}{c|}{0.0309} \\
\multicolumn{1}{c|}{MR} &  & 0.0210 & 0.0642 & 0.0476 & \multicolumn{1}{c|}{0.0573} &  & 0.0195 & 0.0573 & 0.0449 & \multicolumn{1}{c|}{0.0524} &  & 0.0089 & 0.0309 & 0.0170 & \multicolumn{1}{c|}{0.0242} \\
\multicolumn{1}{c|}{CIESS} &  & \textbf{0.0263} & \textbf{0.0730} & \textbf{0.0649} & \multicolumn{1}{c|}{\textbf{0.0705}} &  & \textbf{0.0232} & \textbf{0.0669} & \textbf{0.0551} & \multicolumn{1}{c|}{\textbf{0.0619}} &  & \textbf{0.0153} & \textbf{0.0500} & \textbf{0.0350} & \multicolumn{1}{c|}{\textbf{0.0442}} \\ \hline
\multicolumn{1}{c|}{PEP} & \multirow{4}{*}{95\%} & 0.0199 & 0.0600 & 0.0496 & \multicolumn{1}{c|}{0.0561} & \multirow{4}{*}{95\%} & 0.0075 & 0.0259 & 0.0143 & \multicolumn{1}{c|}{0.0206} & \multirow{4}{*}{95\%} & 0.0125 & 0.0399 & 0.0244 & \multicolumn{1}{c|}{0.0330} \\
\multicolumn{1}{c|}{ES} &  & 0.0217 & 0.0624 & 0.0494 & \multicolumn{1}{c|}{0.0571} &  & 0.0196 & 0.0573 & 0.0496 & \multicolumn{1}{c|}{0.0545} &  & 0.0091 & 0.0303 & 0.0174 & \multicolumn{1}{c|}{0.0238} \\
\multicolumn{1}{c|}{MR} &  & 0.0195 & 0.0583 & 0.0465 & \multicolumn{1}{c|}{0.0540} &  & 0.0175 & 0.0528 & 0.0415 & \multicolumn{1}{c|}{0.0486} &  & 0.0078 & 0.0275 & 0.0145 & \multicolumn{1}{c|}{0.0216} \\
\multicolumn{1}{c|}{CIESS} &  & \textbf{0.0230} & \textbf{0.0657} & \textbf{0.0534} & \multicolumn{1}{c|}{\textbf{0.0640}} &  & \textbf{0.0222} & \textbf{0.0662} & \textbf{0.0540} & \multicolumn{1}{c|}{\textbf{0.0613}} &  & \textbf{0.0157} & \textbf{0.0483} & \textbf{0.0383} & \multicolumn{1}{c|}{\textbf{0.0446}} \\ \hline
\multicolumn{16}{c}{(b) Results on Yelp2018}
\end{tabular}%
}
\caption{Performance of all methods on MovieLens-1M (a) and Yelp2018 (b). R@$k$ and N@$k$ are shorthands for Recall@$k$ and NDCG@$k$, respectively. We highlight the best results when $c$ is set to $80\%$, $90\%$, and $95\%$.
}
\label{tab:overall}
\end{table*}

\subsection{Evaluation Protocols}
We perform evaluation on two popular benchmarks, namely MovieLens-1M \cite{harper2015movielens} with 1,000,208 interactions between 6,040 users and 3,952 movies, and Yelp2018 \cite{wang2019neural} with 1,561,147 interactions between 31,668 users and 38,048 businesses. We split both datasets for training, validation and test with the ratio of 50\%, 25\%, and 25\%. 

For effectiveness, we adopt Recall@$k$ and NDCG@$k$ as our metrics by setting $k\in \{5,20\}$. For CIESS, PEP, ES, and MR, we test the recommendation performance under three sparsity ratios $c\in \{80\%, 90\%, 95\%\}$. For each of these four methods, it is guaranteed that the compressed embedding table has no more than $cd_{max} \times(|\mathcal{U}|+|\mathcal{V}|)$ usable parameters, where the full embedding size $d_{max}=128$. Notably, since ESAPN and OptEmbed have a more performance-oriented design and do not offer a mechanism to precisely control the resulted embedding sparsity, we only report the performance achieved by their final embedding tables. 

\subsection{Implementation Notes for CIESS}
The subsection details our implementation of the proposed model. Both the base recommender and the search function are trained with Adam optimizer \cite{https://doi.org/10.48550/arxiv.1412.6980}. 
We train CIESS for a total of $M=30$ episodes, and each episode contains $T=10$ iterations. The action space is $[1,128]$ for the actor network, and the standard deviation $\sigma = 6$ in the Guassian noise. For the random walk component, we set both the threshold $t$ and walk length to 5. 

\subsection{Comparison of Overall Performance}
Table \ref{tab:overall} shows the performance of all lightweight embedding methods when paired with different base recommenders. In general, at each specified sparsity $c\in \{80\%,90\%,95\%\}$, CIESS significantly outperforms all the precise baselines (i.e., PEP, ES, and MR that have control over the resulted sparsity rate) when paired with all three base recommenders. Specifically, though using a small, fixed embedding size in ES yields competitive recommendation performance compared with PEP's pruning strategy, the embeddings' expressiveness is largely constrained for some important users/items. 

Meanwhile, ESAPN and OptEmbed have respectively resulted in a 75\% and 79\% sparsity rate on average, and failed to meet the lowest 80\% target most of the time. Although both have retained relatively more parameters than all other methods in many test cases, their recommendation performance constantly falls behind CIESS. For example, when paired with LightGCN on MovieLens-1M dataset, ESAPN needs to consume more parameters (72\% embedding sparsity) to obtain a competitive Recall@5 score w.r.t. CIESS under 80\% sparsity. It is also worth noting that, on Yelp2018 dataset, CIESS under 90\% sparsity even outperforms ESAPN and OptEmbed with a much lower sparsity (e.g., 67\% for OptEmbed 76\% for ESAPN). In short, with the same memory consumption, CIESS delivers stronger performance; and at the same performance level, CIESS is more memory-efficient. Hence, the results showcase the continuous embedding size search in CIESS is more advantageous in preserving recommendation accuracy. 

Another interesting finding is that, all methods benefit from a performance increase when paired with a stronger
recommender, especially the graph-based NGCF and its improved variant LightGCN. NCF is generally less accurate
with sparse embeddings, where one possible cause is its matrix factorization component that directly applies dot product to highly sparse user and item embeddings without any deep layers in between. This provides us with some further implications on the choice of base recommenders in a lightweight embedding paradigm.

\begin{table}[t]
\resizebox{0.47\textwidth}{!}{
\begin{tabular}{ccc|cc|cc}
\hline
\multirow{2}{*}{Sparsity} & \multirow{2}{*}{Noise} & Random & \multicolumn{2}{c|}{MovieLens-1M} & \multicolumn{2}{c}{Yelp2018} \\ 
\cline{4-7}
 &  & Walk & R@20 & N@20 & R@20 & N@20 \\ \hline
\multirow{4}{*}{80\%} & G & Yes & \textbf{0.2437} & \textbf{0.4250} & \textbf{0.0839} & \textbf{0.0783} \\
& OU & {Yes} & 0.2432 & 0.4243 & 0.0765 & 0.0722 \\
 & U & Yes & 0.2421 & 0.4241 & 0.0763 & 0.0710 \\
 & G & No & 0.2130 & 0.3825 & 0.0766 & 0.0705 \\ \hline
\multirow{4}{*}{90\%} & G & Yes & \textbf{0.2248} & {\textbf{0.4023}} & 0.0730 & {\textbf{0.0705}} \\
 & {OU} & {Yes} & 0.2212 & 0.3936 & \textbf{0.0734} & 0.0685 \\
 & U & Yes & 0.2212 & {0.3971} & 0.0734 & {0.0692} \\
 & G & No & 0.1995 & 0.3635 & 0.0701 & {0.0644} \\ \hline
\multirow{4}{*}{95\%} & G & Yes & 0.2009 & {\textbf{0.3664}} & \textbf{0.0657} & {\textbf{0.0604}} \\
 & OU & Yes & \textbf{0.2027} & {0.3570} & 0.0610 & {0.0563} \\
 & U & Yes & 0.1972 & 0.3554 & 0.0600 & 0.0554 \\
 & G & No & 0.1890 & 0.3493 & 0.0603 & 0.0551 \\
 \hline
\end{tabular}%
}
\caption{Performance of different CIESS variants. OU, N and U denote noises $\mathcal{N}$ sampled from Ornstein-Uhlenbeck process, Gaussian distribution and uniform distribution, respectively. RW indicates whether random walk is in use.}
\label{tab:variants}
\end{table}

\subsection{Model Component Analysis}
The exploration strategy is crucial for finding optimal embedding sizes in CIESS, where we have proposed a combination of Gaussian noise (Eq.(\ref{eq:actor})) and random walk on the predicted actions. Thus, a natural question is - how useful our random walk-based exploration is, and will a different choice of noise in Eq.(\ref{eq:actor}) substitute its effect? 

To answer this question, we first conduct a quantitative study with three variants of CIESS. The first two variants respectively use a uniform distribution (U) and an Ornstein-Uhlenbeck process (OU) \cite{uhlenbeck1930theory} to replace the Gaussian noise (G) in Eq.(\ref{eq:actor}), while keeping the random walk component. The third variant retains the Gaussian noise but removes the random walk-based exploration. Table \ref{tab:variants} demonstrates the performance change w.r.t. Recall@20 and NDCG@20 in these variants. Due to space limit, we only report results with the best-performing base recommender LightGCN. The results show that, CIESS is relatively insensitive to the choice of noise on the predicted actions, while Gaussian noise obtains better results in most cases. Furthermore, the removal of random walk leads to a significant performance drop, implying the positive contributions from our proposed exploration strategy. 

\begin{figure}[t]
    \centering
    \begin{subfigure}{\linewidth} 
        \centering
        \includegraphics[width=\linewidth]{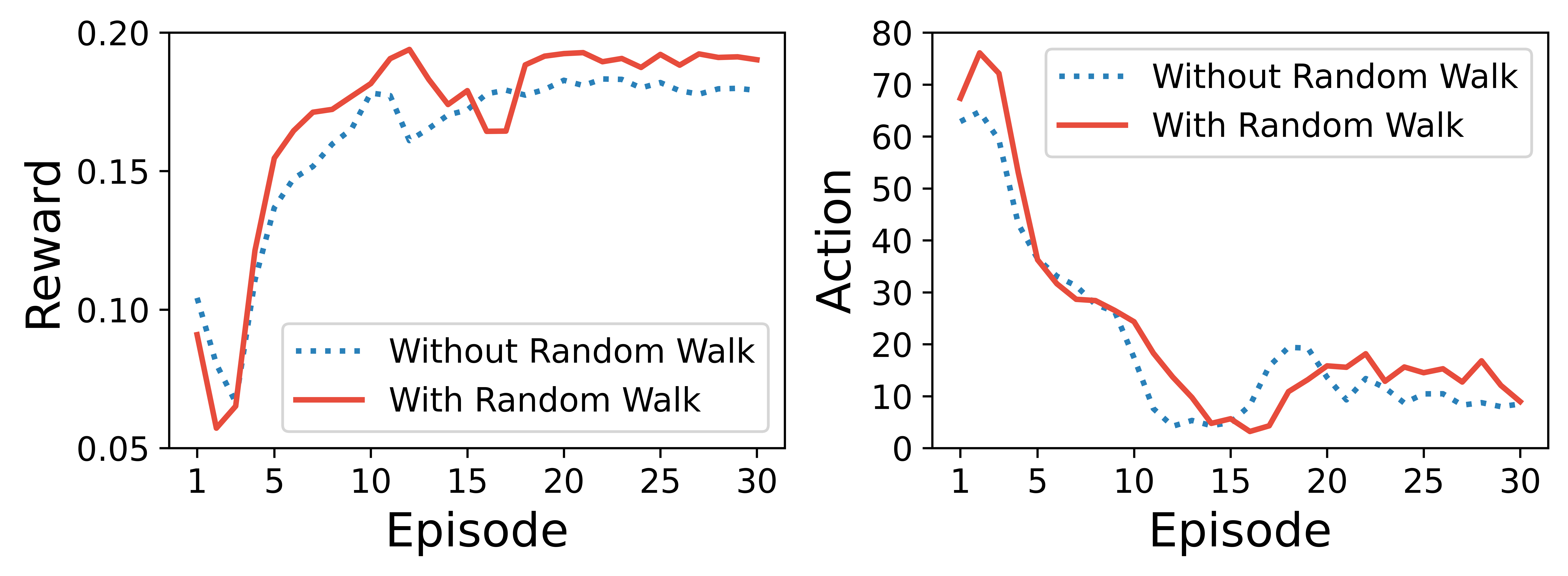}
        \caption{Training Curve on MovieLens-1M}
    \end{subfigure}
    \begin{subfigure}{\linewidth} 
        \centering
        \includegraphics[width=\linewidth]{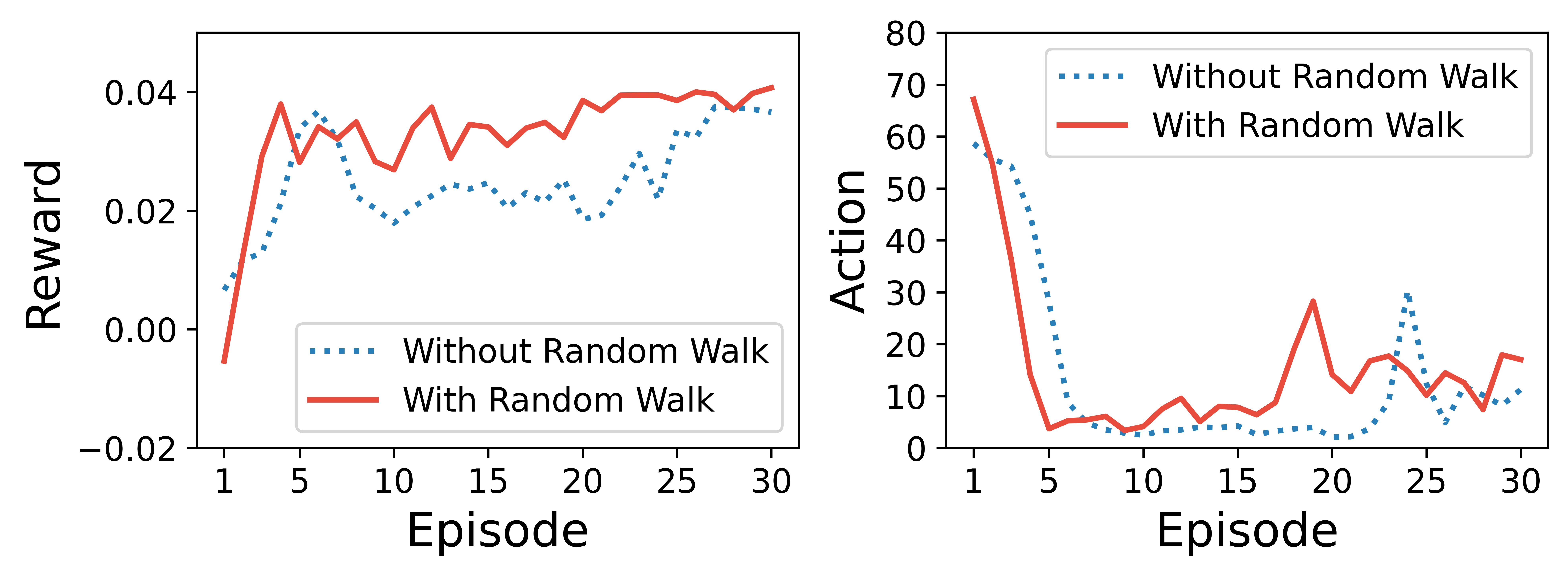} 
        \caption{Training Curve on Yelp2018}
    \end{subfigure}
    \caption{The average reward score and action (embedding size) of CIESS in each training episode on MovieLens-1M (a) and Yelp2018 (b). LightGCN is used as the base recommender.}
    \label{fig:curve}
\end{figure}

Furthermore, we undertake a qualitative analysis by visualizing and comparing the learning processes of CIESS with and without random walk. Figure \ref{fig:curve} displays the average rewards and actions (i.e., embedding sizes) in each episode when both versions are trained. Our observations are: random walk (1) allows the policy to maintain and converge at a higher reward, (2) hit a higher reward sooner, and (3) explores a wider range of actions at both early and final episodes to seek better embedding sizes. 

\subsection{Hyperparameter Sensitivity Analysis}
In this section, we analyze the effect of key hyperparameters in CIESS w.r.t. Recall@20 and NDCG@20. The best base recommender LightGCN is adopted for demonstration.

\subsubsection{Reward Coefficient $\lambda$} 
\begin{figure}[t]
    \begin{subfigure}{\linewidth} 
        \centering
        \includegraphics[width=0.95\linewidth]{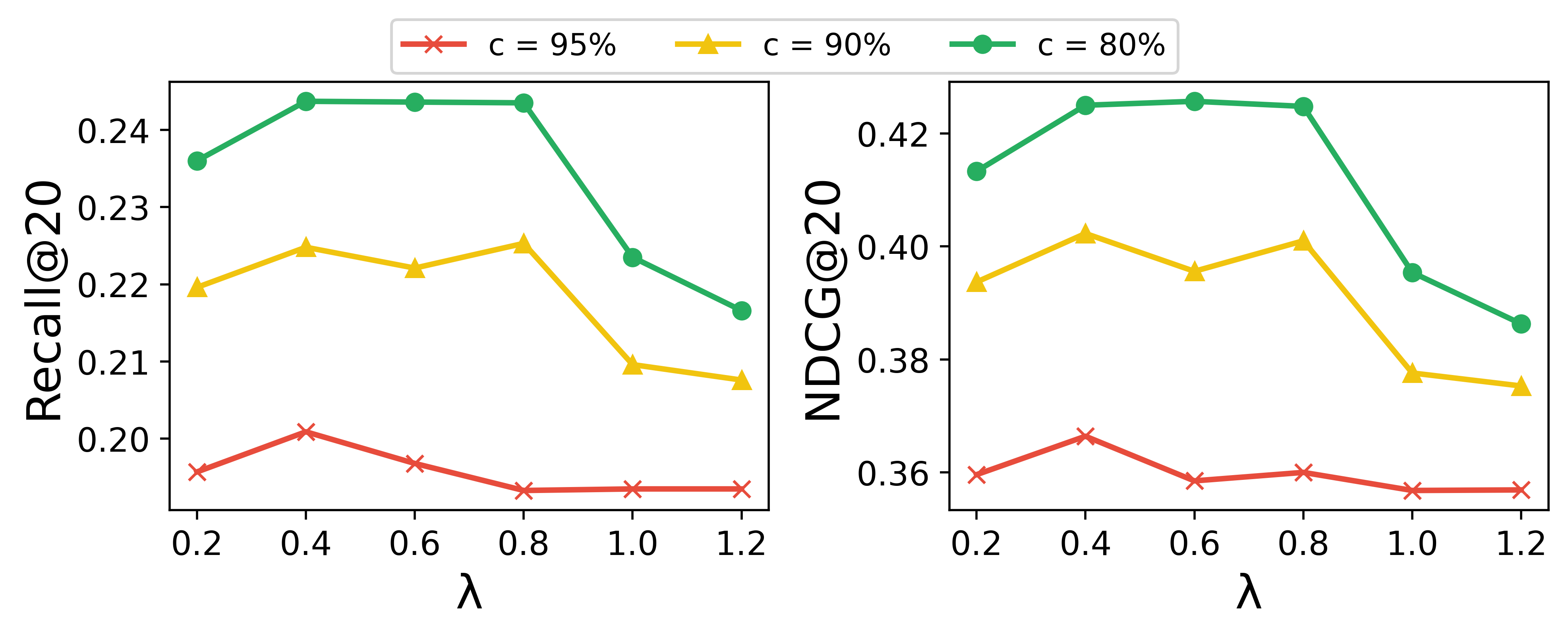} 
        \caption{Effect of $\lambda$ on MovieLens-1M}
    \end{subfigure}
    \begin{subfigure}{\linewidth} 
        \centering
        \includegraphics[width=0.95\linewidth]{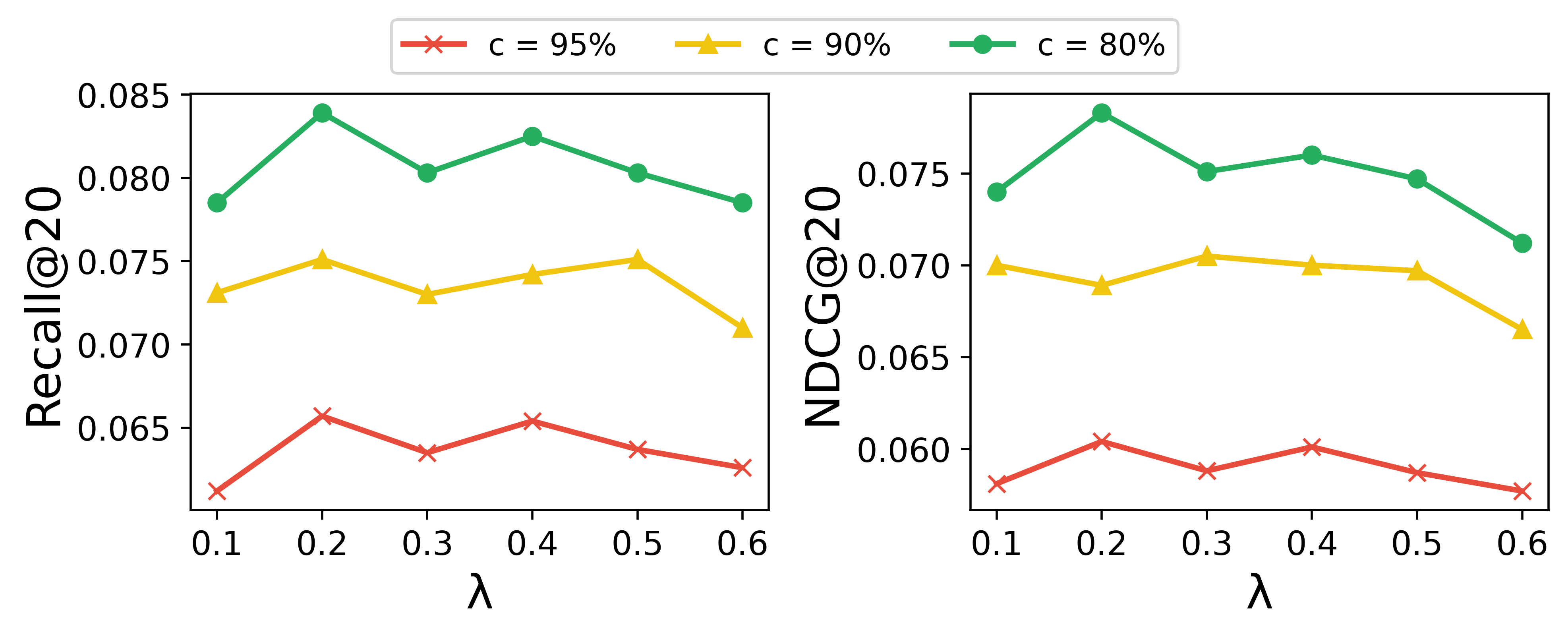} 
        \caption{Effect of $\lambda$ on Yelp2018}
    \end{subfigure}
\caption{Sensitivity analysis w.r.t. $\lambda$. LightGCN is used as the base recommender.} \label{fig:lambda}
\end{figure}

In the reward function Eq.(\ref{eq:reward}), $\lambda$ balances the accuracy and memory consumption objectives. We tune CIESS with $\lambda\in\{0.2, 0.4, 0.6, 0.8, 1.0, 1.2\}$ for MovieLens-1M and $\lambda\in\{0.1, 0.2, 0.3, 0.4, 0.5, 0.6\}$ for Yelp2018, and show how the ranking quality reacts to the change of $\lambda$. As shown in Figure \ref{fig:lambda}(a), CIESS achieves the best performance on MovieLens-1M when $\lambda=0.4$, and the performance starts to deteriorate when $\lambda$ is greater than 0.8. When trained on Yelp2018, Figure \ref{fig:lambda}(b) shows that although the base model performance peaks when $\lambda$ is set to 0.2, it is generally insensitive to the value of $\lambda$ when it ranges between 0.1 and 0.4. After $\lambda$ reaches 0.5, the recommendation performance starts to decline.

\subsubsection{Number of Episodes}
To understand the training efficiency of CIESS, we study how many episodes it needs to reach the best performance. So, we examine its recommendation performance throughout the 30 training episodes by segmenting all episodes into 5 consecutive intervals, and each contains 6 episodes. For each interval, we perform selective retraining described in Section \ref{sec:selec_retrain} for all three target sparsity, and report the best performance observed. 

\begin{figure}[t]
    \begin{subfigure}{\linewidth} 
        \centering
        \includegraphics[width=0.95\linewidth]{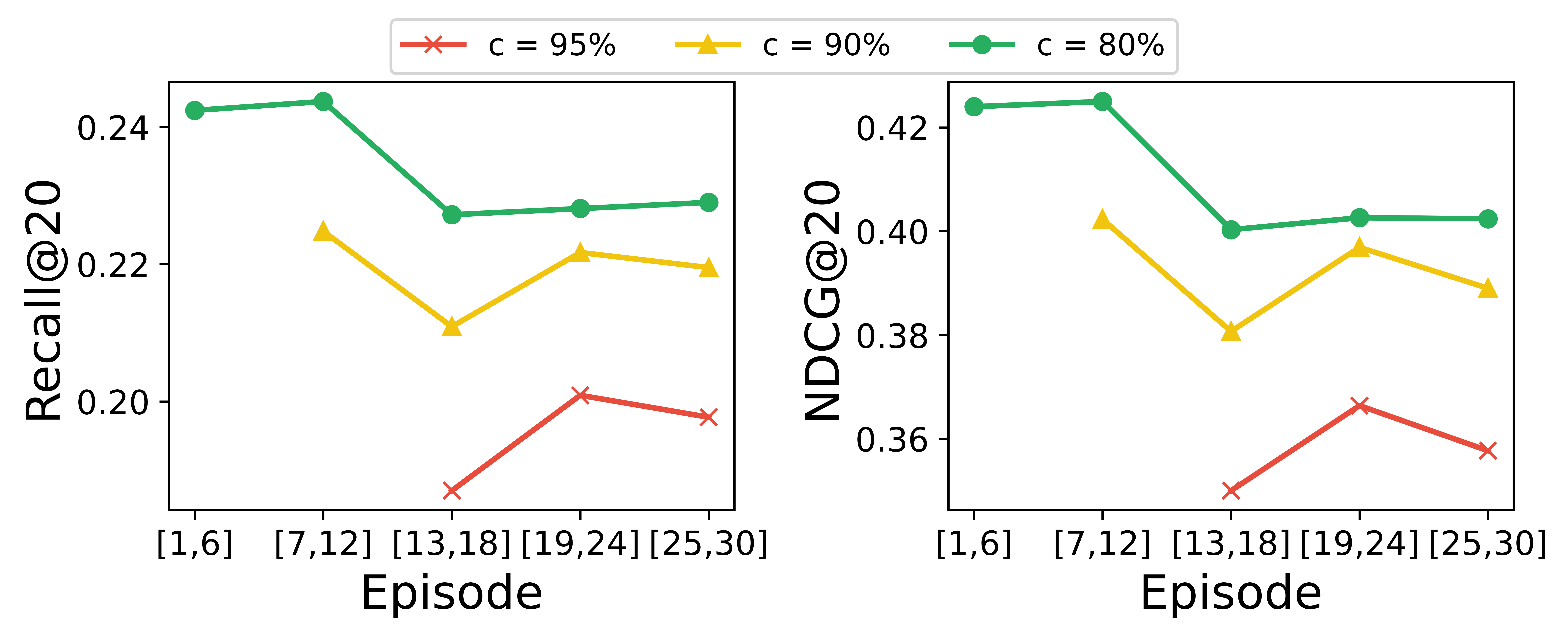} 
        \caption{Effect of $M$ on MovieLens-1M}
    \end{subfigure}
    \begin{subfigure}{\linewidth} 
        \centering
        \includegraphics[width=0.95\linewidth]{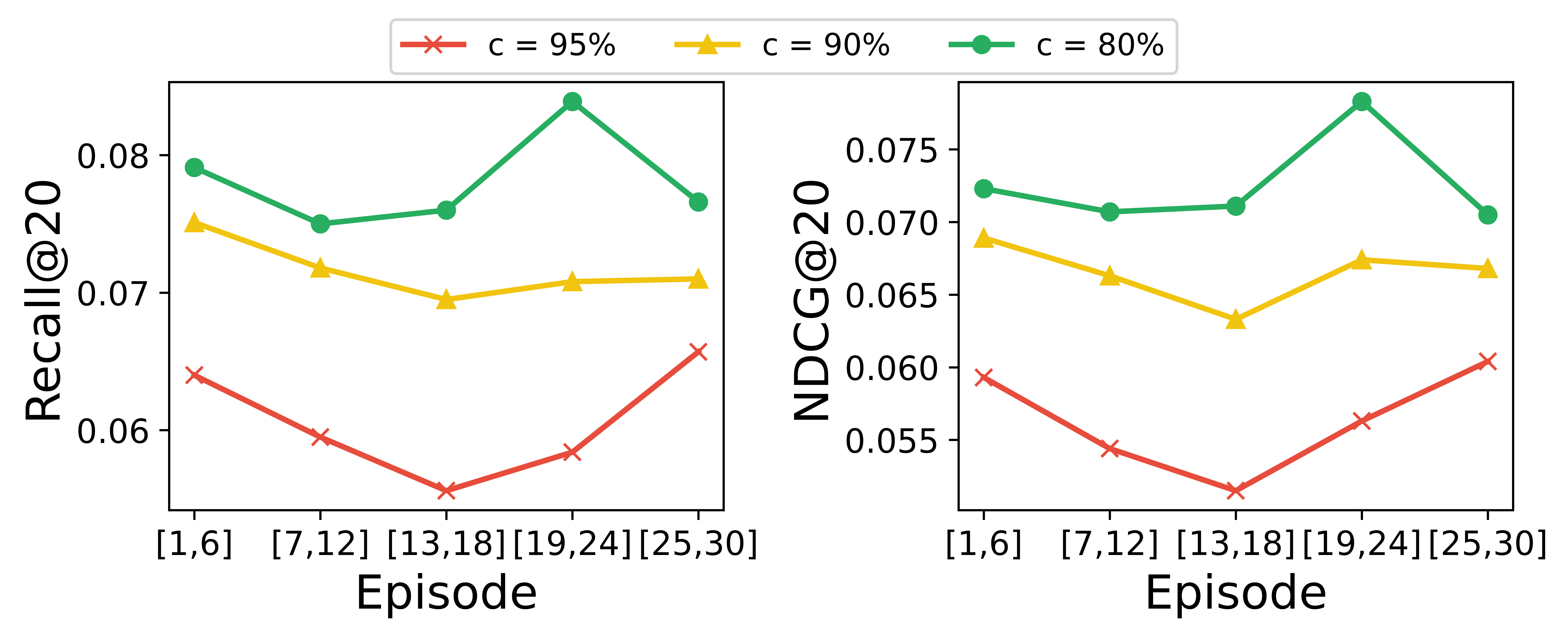} 
        \caption{Effect of $M$ on Yelp2018}
    \end{subfigure}
\caption{Sensitivity analysis w.r.t. $M$. LightGCN is used as the base recommender. $c\in\{90\%,95\%\}$ is reached only in later episodes on MovieLens-1M.}
\label{fig:ab_rw}
\end{figure}

Figure \ref{fig:ab_rw}(a) shows that, when trained on MovieLens-1M, the recommender first reaches its peak performance with $c\in \{80\%, 90\%\}$. The performance of the models within the $c=90\%$ group reaches its height before the 24th episode, and decreases afterwards. On Yelp2018, Figure \ref{fig:ab_rw}(b) indicates that the model performance continues to decline in early episodes when the embedding vectors are being excessively compressed. As the policy gradually compensates the performance by allowing bigger embedding sizes for some users/items, the model performance bounces back and then reaches the highest point in the fourth or fifth episode interval. To summarize, setting $M=30$ is sufficient for CIESS to search for the optimal embedding size when $c$ is high (e.g., 95\%), and it will take less time for a lower $c$ (e.g., 80\%) to optimize.

\section{Related Work}
\subsection{Latent Factor Recommenders} 
Neural networks have exhibited superior ability in solving recommendation tasks. Since MLP can learn the non-linear interactions between users and items, feature representation learning with MLP has been commonplace. He et al. \cite{he2017neural} proposed Neural Collaborative Filtering consisting of an MLP component that learns the non-linear user-item interactions and a generalized matrix factorization component that generalizes the traditional matrix factorization. Cheng et al. \cite{cheng2016wide} proposed Wide\&Deep connecting a wide linear model with a deep neural network for the benefits of memorization and generalization. Apart from MLP-based models, graph-based methods \cite{wang2019neural, he2020lightgcn, graphaug} have also demonstrated promising capability. Wang et al. \cite{wang2019neural} proposed the graph neural network-based model NGCF to model the user-item interactions with a bipartite graph and propagate user-item embeddings on the embedding propagation layers. Similarly, He et al. \cite{he2020lightgcn} resorted to Graph Convolution Network (GCN) and proposed LightGCN that simplified NGCF by including only its neighborhood aggregation component for collaborative filtering. LightGCN performs neighborhood filtering by computing the weighted sum of the user and item embeddings of each layer. Another promising direction, Factorization Machine \cite{rendle2010factorization}, along with its deep variants QNFM \cite{9580543}, DeepFM \cite{guo2017deepfm} and xLightFM \cite{xlightfm} have been studied. \cite{hofm} was also created to mine higher-order interactions. In addition, heterogeneous data such as textual \cite{gong2016hashtag, li2017neural, trythis} and visual \cite{10.1145/3038912.3052638, lee2018collaborative} data has also been exploited by several CNN or RNN-based models. Most of these recommenders utilize vectorized embeddings, which can be optimized by CIESS for memory-efficient embeddings.   

\subsection{AutoML for Recommendation.}
Designing deep recommender models heavily relies on the expertise of professionals. To alleviate the need for human engagement, Automated Machine Learning for Recommender Systems has been studied to automate the design process for recommender models in a task-specific and data-driven manner. So far several research directions including feature selection search \cite{wang2022autofield, luo2019autocross}, embedding dimension search \cite{liu2021learnable, liu2020automated, zhaok2021autoemb, singleshot}, feature interaction search \cite{su2021detecting, su2022detecting}, model architecture search \cite{song2020towards, cheng2022towards}, and other components search \cite{https://doi.org/10.48550/arxiv.2106.06713, meng2021general} have been proposed \cite{automlrecsurvey}. The first kind reduces computation cost by filtering feature fields based on learnable importance scores \cite{wang2022autofield, luo2019autocross}. The second line of works proposes dynamic embedding sizes for each feature \cite{liu2020automated, liu2021learnable, zhaok2021autoemb, singleshot}. The third kind prevents recommender models from enumerating all high-order feature interactions when learning the interactions between features \cite{su2021detecting, su2022detecting}. Model architecture search models explores different network architectures and determines the optimal architecture \cite{song2020towards, cheng2022towards}. Other components search focuses on optimize the loss function and feature interaction function \cite{https://doi.org/10.48550/arxiv.2106.06713, meng2021general}.

CIESS falls into the second group and can derive the optimal embedding sizes in a continuous action space.

\section{Conclusion}
Latent factor recommenders use vectorized embeddings with a single and uniform size, leading to substandard performance and excessive memory complexity. To overcome this issue, we proposed an RL-based, model-agnostic embedding size search algorithm, CIESS, that can select tailored embedding size from a continuous interval for each user/item, thus refining the representation expressiveness with minimal memory costs. Future work could explore richer and more robust importance modeling by incorporating signals such as weight magnitudes \cite{Janowsky1989Pruning, Strom1997Sparse, Thimm1995Evaluating} or model confidence \cite{qu2021human, qu2022combining}. Such extensions could improve capacity allocation for rare-but-informative entities and mitigate popularity bias.

\section*{Acknowledgment} This work is supported by the Australian Research Council under the streams of Future Fellowship (Grant No. FT 210100624), Discovery Project (Grant No. DP190101985), Discovery Early Career Researcher Award (No. DE230101033), and Industrial Transformation Training Centre (No. IC 20010 0022). It is also supported by NSFC (No. 61972069, 61836007 and 61832017), Shenzhen Municipal Science and Technology R\&D Funding Basic Research Program (JCYJ 20210324 133607021), and Municipal Government of Quzhou under Grant No. 2022D037.

\normalem
\bibliographystyle{ACM-Reference-Format}
\balance
\bibliography{ciess}

@article{zhang2019deep,
  title={Deep learning based recommender system: A survey and new perspectives},
  author={Zhang, Shuai and Yao, Lina and Sun, Aixin and Tay, Yi},
  journal={ACM Computing Surveys (CSUR)},
  volume={52},
  number={1},
  pages={1--38},
  year={2019}
}

@article{wang2021survey,
  title={A survey on session-based recommender systems},
  author={Wang, Shoujin and Cao, Longbing and Wang, Yan and Sheng, Quan Z and Orgun, Mehmet A and Lian, Defu},
  journal={ACM Computing Surveys (CSUR)},
  volume={54},
  number={7},
  pages={1--38},
  year={2021}
}

@article{harper2015movielens,
  title={The movielens datasets: History and context},
  author={Harper, F Maxwell and Konstan, Joseph A},
  journal={Acm transactions on interactive intelligent systems (tiis)},
  volume={5},
  number={4},
  pages={1--19},
  year={2015}
}

@inproceedings{liu2020automated,
  title={Automated embedding size search in deep recommender systems},
  author={Liu, Haochen and Zhao, Xiangyu and Wang, Chong and Liu, Xiaobing and Tang, Jiliang},
  booktitle={SIGIR},
  pages={2307--2316},
  year={2020}
}

@inproceedings{zhaok2021autoemb,
  title={Autoemb: Automated embedding dimensionality search in streaming recommendations},
  author={Zhao, Xiangyu and Liu, Haochen and Fan, Wenqi and Liu, Hui and Tang, Jiliang and Wang, Chong and Chen, Ming and Zheng, Xudong and Liu, Xiaobing and Yang, Xiwang},
  booktitle={2021 IEEE International Conference on Data Mining (ICDM)},
  pages={896--905},
  year={2021}
}

@inproceedings{xlightfm,
author = {Jiang, Gangwei and Wang, Hao and Chen, Jin and Wang, Haoyu and Lian, Defu and Chen, Enhong},
title = {XLightFM: Extremely Memory-Efficient Factorization Machine},
year = {2021},
booktitle = {SIGIR},
pages = {337–346},
numpages = {10}
}

@inproceedings{rendle_bpr_2012,
author = {Rendle, Steffen and Freudenthaler, Christoph and Gantner, Zeno and Schmidt-Thieme, Lars},
title = {BPR: Bayesian Personalized Ranking from Implicit Feedback},
year = {2009},
booktitle = {Proceedings of the Twenty-Fifth Conference on Uncertainty in Artificial Intelligence},
pages = {452–461},
year={2009}
}

@inproceedings{ddpg,
  author    = {Timothy P. Lillicrap and
               Jonathan J. Hunt and
               Alexander Pritzel and
               Nicolas Heess and
               Tom Erez and
               Yuval Tassa and
               David Silver and
               Daan Wierstra},
  title     = {Continuous control with deep reinforcement learning},
  booktitle = {ICLR},
  year      = {2016}
}

@article{uhlenbeck1930theory,
  title = {On the Theory of the Brownian Motion},
  author = {Uhlenbeck, G. E. and Ornstein, L. S.},
  journal = {Physical Review Journals},
  volume = {36},
  pages = {823--841},
  year = {1930}
}

@article{dulac2015deep,
  title={Deep reinforcement learning in large discrete action spaces},
  author={Dulac-Arnold, Gabriel and Evans, Richard and van Hasselt, Hado and Sunehag, Peter and Lillicrap, Timothy and Hunt, Jonathan and Mann, Timothy and Weber, Theophane and Degris, Thomas and Coppin, Ben},
  journal={arXiv preprint arXiv:1512.07679},
  year={2015}
}

@inproceedings{joglekar2020neural,
  title={Neural input search for large scale recommendation models},
  author={Joglekar, Manas R and Li, Cong and Chen, Mei and Xu, Taibai and Wang, Xiaoming and Adams, Jay K and Khaitan, Pranav and Liu, Jiahui and Le, Quoc V},
  booktitle={SIGKDD},
  pages={2387--2397},
  year={2020}
}

@inproceedings{sedaghati2015automatic,
  title={Automatic selection of sparse matrix representation on GPUs},
  author={Sedaghati, Naser and Mu, Te and Pouchet, Louis-Noel and Parthasarathy, Srinivasan and Sadayappan, P},
  booktitle={ACM International Conference on Supercomputing},
  pages={99--108},
  year={2015}
}

@article{virtanen2020scipy,
  title={SciPy 1.0: fundamental algorithms for scientific computing in Python},
  author={Virtanen, Pauli and others},
  journal={Nature Methods},
  volume={17},
  number={3},
  pages={261--272},
  year={2020}
}

@inproceedings{liu2021learnable,
title={Learnable Embedding sizes for Recommender Systems},
author={Siyi Liu and Chen Gao and Yihong Chen and Depeng Jin and Yong Li},
booktitle={ICLR},
year={2021}
}

@inproceedings{he2020lightgcn,
  title={Lightgcn: Simplifying and powering graph convolution network for recommendation},
  author={He, Xiangnan and Deng, Kuan and Wang, Xiang and Li, Yan and Zhang, Yongdong and Wang, Meng},
  booktitle={SIGIR},
  pages={639--648},
  year={2020}
}

@inproceedings{he2017neural,
  title={Neural collaborative filtering},
  author={He, Xiangnan and Liao, Lizi and Zhang, Hanwang and Nie, Liqiang and Hu, Xia and Chua, Tat-Seng},
  booktitle={WWW},
  pages={173--182},
  year={2017}
}

@inproceedings{wang2019neural,
  title={Neural graph collaborative filtering},
  author={Wang, Xiang and He, Xiangnan and Wang, Meng and Feng, Fuli and Chua, Tat-Seng},
  booktitle={SIGIR},
  pages={165--174},
  year={2019}
}

@inproceedings{cheng2016wide,
  title={Wide \& deep learning for recommender systems},
  author={Cheng, Heng-Tze and Koc, Levent and Harmsen, Jeremiah and Shaked, Tal and Chandra, Tushar and Aradhye, Hrishi and Anderson, Glen and Corrado, Greg and Chai, Wei and Ispir, Mustafa and others},
  booktitle={Proceedings of the 1st workshop on deep learning for recommender systems},
  pages={7--10},
  year={2016}
}

@inproceedings{rendle2010factorization,
  title={Factorization machines},
  author={Rendle, Steffen},
  booktitle={2010 IEEE International conference on data mining},
  pages={995--1000},
  year={2010}
}

@article{guo2017deepfm,
  title={DeepFM: a factorization-machine based neural network for CTR prediction},
  author={Guo, Huifeng and Tang, Ruiming and Ye, Yunming and Li, Zhenguo and He, Xiuqiang},
  journal={IJCAI},
  year={2017}
}

@inproceedings{lian2020lightrec,
  title={Lightrec: A memory and search-efficient recommender system},
  author={Lian, Defu and Wang, Haoyu and Liu, Zheng and Lian, Jianxun and Chen, Enhong and Xie, Xing},
  booktitle={WWW},
  pages={695--705},
  year={2020}
}

@inproceedings{td3,
    title={Addressing Function Approximation Error in Actor-Critic Methods},
    author={Fujimoto, Scott and van Hoof, Herke and Meger, David},
    booktitle={Proceedings of the 35th International Conference on Machine Learning},
    pages={1587--1596},
    year={2018},
    volume={80}
}

@article{liu2023actor,
  title={Actor-Director-Critic: A Novel Deep Reinforcement Learning Framework},
  author={Liu, Zongwei and Song, Yonghong and Zhang, Yuanlin},
  journal={arXiv preprint arXiv:2301.03887},
  year={2023}
}

@inproceedings{optemb,
    author = {Lyu, Fuyuan and Tang, Xing and Zhu, Hong and Guo, Huifeng and Zhang, Yingxue and Tang, Ruiming and Liu, Xue},
    title = {OptEmbed: Learning Optimal Embedding Table for Click-through Rate Prediction},
    year = {2022},
    booktitle = {CIKM},
    pages = {1399–1409}
}

@article{automlrecsurvey,
    title={AutoML for Deep Recommender Systems: A Survey},
    author={Ruiqi Zheng and Liang Qu and Bin Cui and Yuhui Shi and Hongzhi Yin},
    journal={ACM Transactions on Information Systems},
    year={2022},
    volume = {41},
    number = {4}
}

@inproceedings{hofm,
author = {Blondel, Mathieu and Fujino, Akinori and Ueda, Naonori and Ishihata, Masakazu},
title = {Higher-Order Factorization Machines},
year = {2016},
booktitle = {Proceedings of the 30th International Conference on Neural Information Processing Systems},
pages = {3359–3367}
}

@inproceedings{gong2016hashtag,
  title={Hashtag recommendation using attention-based convolutional neural network.},
  author={Gong, Yuyun and Zhang, Qi},
  booktitle={IJCAI},
  pages={2782--2788},
  year={2016}
}

@inproceedings{li2017neural,
  title={Neural rating regression with abstractive tips generation for recommendation},
  author={Li, Piji and Wang, Zihao and Ren, Zhaochun and Bing, Lidong and Lam, Wai},
  booktitle={Proceedings of the 40th International ACM SIGIR conference on Research and Development in Information Retrieval},
  pages={345--354},
  year={2017}
}

@inproceedings{10.1145/3038912.3052638,
author = {Wang, Suhang and Wang, Yilin and Tang, Jiliang and Shu, Kai and Ranganath, Suhas and Liu, Huan},
title = {What Your Images Reveal: Exploiting Visual Contents for Point-of-Interest Recommendation},
year = {2017},
booktitle = {Proceedings of the 26th International Conference on World Wide Web},
pages = {391–400}
}

@inproceedings{lee2018collaborative,
  title={Collaborative deep metric learning for video understanding},
  author={Lee, Joonseok and Abu-El-Haija, Sami and Varadarajan, Balakrishnan and Natsev, Apostol},
  booktitle={Proceedings of the 24th ACM SIGKDD International conference on knowledge discovery \& data mining},
  pages={481--490},
  year={2018}
}

@inproceedings{wang2022autofield,
  title={Autofield: Automating feature selection in deep recommender systems},
  author={Wang, Yejing and Zhao, Xiangyu and Xu, Tong and Wu, Xian},
  booktitle={WWW},
  pages={1977--1986},
  year={2022}
}

@inproceedings{luo2019autocross,
  title={Autocross: Automatic feature crossing for tabular data in real-world applications},
  author={Luo, Yuanfei and Wang, Mengshuo and Zhou, Hao and Yao, Quanming and Tu, Wei-Wei and Chen, Yuqiang and Dai, Wenyuan and Yang, Qiang},
  booktitle={SIGKDD},
  pages={1936--1945},
  year={2019}
}

@inproceedings{su2021detecting,
  title={Detecting beneficial feature interactions for recommender systems},
  author={Su, Yixin and Zhang, Rui and Erfani, Sarah and Xu, Zhenghua},
  booktitle={AAAI},
  number={5},
  pages={4357--4365},
  year={2021}
}

@inproceedings{su2022detecting,
  title={Detecting arbitrary order beneficial feature interactions for recommender systems},
  author={Su, Yixin and Zhao, Yunxiang and Erfani, Sarah and Gan, Junhao and Zhang, Rui},
  booktitle={SIGKDD},
  pages={1676--1686},
  year={2022}
}

@inproceedings{cheng2022towards,
  title={Towards Automatic Discovering of Deep Hybrid Network Architecture for Sequential Recommendation},
  author={Cheng, Mingyue and Liu, Zhiding and Liu, Qi and Ge, Shenyang and Chen, Enhong},
  booktitle={WWW},
  pages={1923--1932},
  year={2022}
}

@inproceedings{song2020towards,
  title={Towards automated neural interaction discovery for click-through rate prediction},
  author={Song, Qingquan and Cheng, Dehua and Zhou, Hanning and Yang, Jiyan and Tian, Yuandong and Hu, Xia},
  booktitle={SIGKDD},
  pages={945--955},
  year={2020}
}

@inproceedings{https://doi.org/10.48550/arxiv.2106.06713,
author = {Zhao, Xiangyu and Liu, Haochen and Fan, Wenqi and Liu, Hui and Tang, Jiliang and Wang, Chong},
title = {AutoLoss: Automated Loss Function Search in Recommendations},
year = {2021},
booktitle = {SIGKDD},
pages = {3959–3967}
}

@inproceedings{meng2021general,
  title={A general method for automatic discovery of powerful interactions in click-through rate prediction},
  author={Meng, Ze and Zhang, Jinnian and Li, Yumeng and Li, Jiancheng and Zhu, Tanchao and Sun, Lifeng},
  booktitle={SIGIR},
  pages={1298--1307},
  year={2021}
}

@inproceedings{10.1145/3366423.3380170,
author = {Wang, Qinyong and Yin, Hongzhi and Chen, Tong and Huang, Zi and Wang, Hao and Zhao, Yanchang and Viet Hung, Nguyen Quoc},
title = {Next Point-of-Interest Recommendation on Resource-Constrained Mobile Devices},
year = {2020},
booktitle = {WWW},
pages = {906–916}
}

@inproceedings{10.1145/3447548.3467220,
author = {Chen, Tong and Yin, Hongzhi and Zheng, Yujia and Huang, Zi and Wang, Yang and Wang, Meng},
title = {Learning Elastic Embeddings for Customizing On-Device Recommenders},
year = {2021},
booktitle = {SIGKDD},
pages = {138–147}
}

@inproceedings{shi2020compositional,
  title={Compositional embeddings using complementary partitions for memory-efficient recommendation systems},
  author={Shi, Hao-Jun Michael and Mudigere, Dheevatsa and Naumov, Maxim and Yang, Jiyan},
  booktitle={SIGKDD},
  pages={165--175},
  year={2020}
}

@inproceedings{10.1145/2911451.2911502,
author = {Zhang, Hanwang and Shen, Fumin and Liu, Wei and He, Xiangnan and Luan, Huanbo and Chua, Tat-Seng},
title = {Discrete Collaborative Filtering},
year = {2016},
booktitle = {SIGIR},
pages = {325–334}
}

@inproceedings{https://doi.org/10.48550/arxiv.2010.10784,
    author = {Kang, Wang-Cheng and Cheng, Derek Zhiyuan and Yao, Tiansheng and Yi, Xinyang and Chen, Ting and Hong, Lichan and Chi, Ed H.},
    title = {Learning to Embed Categorical Features without Embedding Tables for Recommendation},
    year = {2021},
    pages = {840–850},
    booktitle = {SIGKDD}
}

@inproceedings{10.1145/3477495.3531775,
author = {Xia, Xin and Yin, Hongzhi and Yu, Junliang and Wang, Qinyong and Xu, Guandong and Nguyen, Quoc Viet Hung},
title = {On-Device Next-Item Recommendation with Self-Supervised Knowledge Distillation},
year = {2022},
booktitle = {SIGIR},
pages = {546–555}
}

@article{https://doi.org/10.48550/arxiv.1809.02121,
  author = {Zahavy, Tom and Haroush, Matan and Merlis, Nadav and Mankowitz, Daniel J. and Mannor, Shie},
  title = {Learn What Not to Learn: Action Elimination with Deep Reinforcement Learning},
  year = {2018},
  booktitle = {NeurIPS},
  pages = {3566–3577}
}

@article{grondman2012survey,
  title={A survey of actor-critic reinforcement learning: Standard and natural policy gradients},
  author={Grondman, Ivo and Busoniu, Lucian and Lopes, Gabriel AD and Babuska, Robert},
  journal={IEEE Transactions on Systems, Man, and Cybernetics, Part C (Applications and Reviews)},
  volume={42},
  number={6},
  pages={1291--1307},
  year={2012}
}

@inproceedings{https://doi.org/10.48550/arxiv.1412.6980,
  author    = {Diederik P. Kingma and Jimmy Ba},
  editor    = {Yoshua Bengio and Yann LeCun},
  title     = {Adam: {A} Method for Stochastic Optimization},
  booktitle = {ICLR},
  year      = {2015},
}

@article{krichene2022sampled,
  title={On sampled metrics for item recommendation},
  author={Krichene, Walid and Rendle, Steffen},
  journal={Communications of the ACM},
  volume={65},
  number={7},
  pages={75--83},
  year={2022}
}

@inproceedings{li2021lightweight,
  title={Lightweight self-attentive sequential recommendation},
  author={Li, Yang and Chen, Tong and Zhang, Peng-Fei and Yin, Hongzhi},
  booktitle={CIKM},
  pages={967--977},
  year={2021}
}

@article{9580543,
  author={Chen, Tong and Yin, Hongzhi and Zhang, Xiangliang and Huang, Zi and Wang, Yang and Wang, Meng},
  journal={IEEE Transactions on Neural Networks and Learning Systems}, 
  title={Quaternion Factorization Machines: A Lightweight Solution to Intricate Feature Interaction Modeling}, 
  year={2021},
  volume={},
  number={},
  pages={1-14},
  doi={10.1109/TNNLS.2021.3118706}
}

@inproceedings{trythis,
author = {Chen, Tong and Yin, Hongzhi and Ye, Guanhua and Huang, Zi and Wang, Yang and Wang, Meng},
title = {Try This Instead: Personalized and Interpretable Substitute Recommendation},
year = {2020},
booktitle = {SIGIR},
pages = {891–900}
}

@article{10.1145/3580364,
author = {Xia, Xin and Yu, Junliang and Wang, Qinyong and Yang, Chaoqun and Hung, Nguyen Quoc Viet and Yin, Hongzhi},
title = {Efficient On-Device Session-Based Recommendation},
year = {2023},
volume = {41},
number = {4},
journal = {ACM Trans. Inf. Syst.}
}

@inproceedings{singleshot,
author = {Qu, Liang and Ye, Yonghong and Tang, Ningzhi and Zhang, Lixin and Shi, Yuhui and Yin, Hongzhi},
title = {Single-Shot Embedding Dimension Search in Recommender System},
year = {2022},
booktitle = {SIGIR},
pages = {513–522},
}

@inproceedings{graphaug,
author = {Yu, Junliang and Yin, Hongzhi and Xia, Xin and Chen, Tong and Cui, Lizhen and Nguyen, Quoc Viet Hung},
title = {Are Graph Augmentations Necessary? Simple Graph Contrastive Learning for Recommendation},
year = {2022},
pages = {1294–1303},
booktitle = {SIGIR}
}

@article{qu2022combining,
    author = {Qu, Yunke and Roitero, Kevin and Barbera, David La and Spina, Damiano and Mizzaro, Stefano and Demartini, Gianluca},
    title = {Combining Human and Machine Confidence in Truthfulness Assessment},
    year = {2022},
    volume = {15},
    number = {1},
    journal = {J. Data and Information Quality}
}

@inproceedings{qu2021human,
    title={Human-in-the-Loop Systems for Truthfulness: A Study of Human and Machine Confidence},
    author={Yunke Qu and Kevin Roitero and Stefano Mizzaro and Damiano Spina and Gianluca Demartini},
    booktitle={Conference for Truth and Trust Online},
    year={2021}
}

@article{Janowsky1989Pruning,
    title = {Pruning versus clipping in neural networks},
    author = {Janowsky, Steven A.},
    journal = {Phys. Rev. A},
    volume = {39},
    pages = {6600--6603},
    year = {1989},
    number = {}
}

@inproceedings{Strom1997Sparse,
    author={Nikko Ström},
    title={{Sparse connection and pruning in large dynamic artificial neural networks}},
    year={1997},
    booktitle={Proc. 5th European Conf. on Speech Communication and Technology},
    pages={2807--2810}
}

@inproceedings{Thimm1995Evaluating,
    title={Evaluating pruning methods},
    author={Thimm, Georg and Fiesler, Emile},
    booktitle={Proc. of the Intl. Symposium on Artificial neural networks},
    pages={20--25},
    year={1995}
}

\end{document}